\documentclass[fleqn,usenatbib]{mnras}

\usepackage{newtxtext,newtxmath}

\usepackage[T1]{fontenc}
\usepackage{ae,aecompl}

\usepackage{graphicx}	
\usepackage{amsmath}	
\usepackage{amssymb}	

\newcommand{\Nstarlong}{NGTS J093016-185033.6}
\newcommand{\Nstar}{NGTS J0930-18}

\newcommand{\feh}{[Fe/H]}
\newcommand{\teff}{$\mathrm{T_{eff}}$}
\newcommand{\logg}{$\log g$}
\newcommand{\masy}{mas\,yr$^{-1}$}

\newcommand{\massratio}{$0.1407 ^{+0.0065}_{-0.017}$}
\newcommand{\period}{$1.33265 $}
\newcommand{\ecc}{$0.00592 ^{+0.0098}_{-0.0035}$}

\newcommand{\PristarRad}{$0.584 ^{+0.0094}_{-0.010}$} 
\newcommand{\PristarMass}{$0.5803 ^{+0.0092}_{-0.0063}$} 
\newcommand{\NstarRad}{$0.1059 ^{+0.0023}_{-0.0021}$} 
\newcommand{\NstarRadJup}{$1.052 ^{+0.023}_{-0.021}$} 
\newcommand{\NstarMass}{$0.0818 ^{+0.0040}_{-0.0015}$} 
\newcommand{\NstarMassJup}{$85.7 ^{+4.2}_{-1.5}$} 

\title[Unequal Mass M-dwarf binary]{An eclipsing M-dwarf close to the hydrogen burning limit from NGTS}

\author[J. S. Acton et al.]{
\parbox{\textwidth}{
Jack S.~Acton,$^{1}$\thanks{E-mail:ja466@le.ac.uk}
Michael R.~Goad,$^{1}$
Sarah L.~Casewell,$^{1}$
Jose I. Vines,$^{2}$
Matthew R.~Burleigh,$^{1}$
Philipp Eigm\"uller, $^{3}$
Louise D. Nielsen,$^{4}$
Boris T. G\"ansicke,$^{5}$
Daniel Bayliss $^{5}$
Fran\c{c}ois Bouchy,$^{4}$ 
Edward M.~Bryant,$^{5}$
Samuel Gill,$^{5}$
Edward Gillen,$^{6,7}$\thanks{Winton Fellow}
Maximilian N. G{\"u}nther,$^{8}$\thanks{Juan Carlos Torres Fellow}
James S. Jenkins,$^{2,9}$
James McCormac,$^{5}$
Maximiliano Moyano,$^{10}$
Liam R.~Raynard,$^{1}$
Rosanna H.~Tilbrook,$^{1}$
St\'ephane Udry,$^{4}$
Christopher A. Watson,$^{11}$
Richard G.\ West,$^{5}$
Peter J.\ Wheatley,$^{5}$
}
\\
$^{1}$School of Physics and Astronomy, University of Leicester, University Road, Leicester, LE1 7RH, UK\\
$^{2}$Departamento de Astronom\'ia, Universidad de Chile, Camino el Observatorio 1515, Las Condes, Santiago, Chile\\
$^{3}$Institute of Planetary Research, German Aerospace Center, Rutherfordstrasse 2, 12489, Berlin, Germany\\
$^{4}$Geneva Observatory, University of Geneva, Chemin des Mailettes 51, 1290 Versoix, Switzerland\\
$^{5}$Department of Physics, University of Warwick, Gibbet Hill Road, Coventry CV4 7AL, UK\\
$^{6}$Astronomy Unit, Queen Mary University of London, Mile End Road, London E1 4NS, UK \\
$^{7}$Astrophysics Group, Cavendish Laboratory, J.J. Thomson Avenue, Cambridge CB3 0HE, UK\\
$^{8}$Department of Physics, and Kavli Institute for Astrophysics and Space Research, Massachusetts Institute of Technology, Cambridge, MA 02139, USA\\
$^{9}$Centro de Astrof\'isica y Tecnolog\'ias Afines (CATA), Casilla 36-D, Santiago, Chile\\
$^{10}$Instituto de Astronom\'ia, Universidad Cat\'olica del Norte, Angamos 0610, 1270709, Antofagasta, Chile\\
$^{11}$ Astrophysics Research Centre, School of Mathematics and Physics, Queen’s University Belfast, Belfast, BT7 1NN, UK\\
}

\date{Accepted XXX. Received YYY; in original form ZZZ}

\pubyear{2020}
\hypersetup{draft}
\begin{document}
\label{firstpage}
\pagerange{\pageref{firstpage}--\pageref{lastpage}}
\maketitle

\begin{abstract}
We present the discovery of \Nstar, an extreme mass ratio eclipsing M-dwarf binary system with an early M-dwarf primary and a late M-dwarf secondary close to the hydrogen burning limit. 
Global modelling of photometry and radial velocities reveals that the secondary component (\Nstar\,B) has a mass of M$_*$=\NstarMass\ M$_{\odot}$ and radius of R$_*$=\NstarRad\ R$_{\odot}$, making it one of the lowest mass stars with direct mass and radius measurements. With a mass ratio of q = \massratio, \Nstar\ has the lowest mass ratio of any known eclipsing M-dwarf binary system, posing interesting questions for binary star formation and evolution models. The mass and radius of \Nstar\,B is broadly consistent with stellar evolutionary models. \Nstar\,B lies in the sparsely populated mass radius parameter space close to the substellar boundary. Precise measurements of masses and radii from single lined eclipsing binary systems of this type are vital for constraining the uncertainty in the mass-radius relationship - of importance due to the growing number of terrestrial planets being discovered around low mass stars.
\end{abstract}




\section{Introduction}
Eclipsing binary stars are of vital importance in the field of stellar structure.  These are the only objects from which we are able to get accurate mass--radius measurements of stars to test against model predictions. This is particularly relevant for low mass stars (known as M-dwarfs). Stars with masses below 0.25$M_\odot$ are the most common stellar objects \citep{Henry2006} but despite this the physics governing them remains relatively poorly understood. In particular, the mass--radius relation for low mass stars is poorly constrained when compared with theoretical models \citep{Parsons2018}.

Previous studies have shown that model predictions for masses and radii of M-dwarfs can differ from measured values by up to 10 per cent (\citealt{Feiden2012, Terrien2012}). This discrepancy is most likely due to magnetic activity induced by interactions in short-period tidally-locked binary systems (\citealt{Ribas2006, Chabrier2007}). However, this is not entirely clear due to the existence of longer period systems, which we would not expect to be tidally locked, that show the same over sizing as these short period systems (\citealt{Doyle2011}, \citealt{Irwin2011}), as well as short period systems that show good agreement with models (\citealt{Blake2008}). This is further complicated by the fact that it is expected that fully convective stars (those with masses $<$ 0.35 $M_{\odot}$) should show less inflation due to the nature of their atmospheres \citep{Kraus2011}. 

It is then vital that we are able to further constrain models for low mass stars by obtaining direct measurements of stellar masses and radii. This has motivated the search for examples of low-mass eclipsing binaries (also known as EBLMs) that have provided accurate mass and radius measurements for a large number of M-dwarf stars across a range of masses (e.g \citealt{Triaud2012, Chew2014, von_Boetticher2017, Triaud2017, vonBoetticher_2019, Gill2019b}). However the lowest masses, between 0.1 $M_{\odot}$ and hydrogen burning limit at $\sim 0.07 M_{\odot}$, remains relatively sparsely sampled. 

When characterising exoplanet systems, accurate knowledge of the host star's parameters is crucial as these are used to determine the corresponding values for the planet. Uncertainties in the stellar values could lead to over or underestimation of discovered planetary masses and radii. This is of further importance as some of the most interesting planetary systems have been discovered around low mass stars (e.g., \citealt{Gillon2017, Guenther2019, Kostov2019}). The smaller radii of these stars means that small planets produce transits of a much greater depth than the same planet occulting a larger star, making them much easier to detect in transit surveys. For this reason many modern transiting exoplanet surveys are designed to target such low mass stars (e.g., TRAPPIST \citep{Gillon2011}, SPECULOOS \citep{Delrez2018}).

Additionally, observations of rare types of eclipsing binary systems can provide insights into star formation. For example, \citealt{Wisniewski2012} proposed that there should be a lack of binaries with orbital periods less than 100 days with highly unequal mass components. There are examples of so called Extreme Mass Ratio Binaries (EMRBs) comprising a large star (typically A or B spectral type) with an M-dwarf companion (e.g., \citealt{Stevens2019}), however the vast majority of eclipsing M-dwarf binaries are systems of roughly equal mass \citep{Delfosse2004}. \citealt{Laithwaite2020} surveyed a large sample of late M-dwarf binaries and found that they are almost exclusively equal mass systems. This is possibly a formation effect.  \citealt{Bouchy2011} propose that stars with spectral types later than G have disk braking strong enough to cause low mass short period companions to migrate inwards and become engulfed. Therefore we would expect extreme mass ratio binaries of two low mass stars to be rare.

In this paper we present the discovery of \Nstarlong\ (hereafter \Nstar), a highly unequal mass ratio (q=0.14075) eclipsing M-dwarf binary in which the secondary (\Nstar\,B) is a very low mass star just above the classical hydrogen burning limit of $\sim$ 70\,M\textsubscript{J} \citep{Dieterich2014}. We make use of follow-up photometric and spectroscopic observations to determine accurate masses and radii for the star, which lies in a region of parameter space with few direct measurements. This discovery will aid in further constraining the lowest end of the stellar mass-radius relationship.

\section{Observations}
\begin{table*}
	\centering
	\caption{Summary of observations.}
	\label{tab:obs_summary}
	\begin{tabular}{ccccccc} 
    \hline
Observation type & Telescope & Band  & Cadence & Total integration time & Period & Notes\\
\hline
Photometry	& NGTS	&	520-890\,nm	&	13\,s   & 156\,nights	    &	21/04/16-22/12/16 & 14 full eclipses\\
Photometry	& SAAO	&	I	        &	30\,s   &3\,hours  	    &	20/12/18 & Single Observation\\
Photometry	& SAAO	&	g'	        &	10\,s   & 2.16\,hours	    &	29/01/19 & Single Observation\\
Photometry	& TESS	&  600-1000\,nm	&	1800\,s & 28\,days	    &	02/02/19-27/02/19 & 10 eclipses in total\\
Spectroscopy& HARPS	&	378-691\,nm	        &      45\,minutes &	      4.5\,hours	    &   11/04/19-08/06/19          & Six RV Points, EGGS mode \\

	\hline
    \end{tabular}
\end{table*}

\Nstar\ was initially discovered using photometry from NGTS \citep{Wheatley2018}. Follow-up observations were performed with the Sutherland High Speed Optical Cameras (SHOC) \citep{Coppejans2013} on the South African Astronomical Observatory (SAAO) 1-m telescope. This photometry was then used in conjunction with observations from the Transiting Exoplanet Survey Satellite (TESS, \citealt{Ricker2014}). We obtained high resolution spectra with HARPS (mounted on the ESO 3.6m, \citealt{Mayor2003}) to determine the mass of the companion. These observations are detailed in Table \ref{tab:obs_summary} and described below.

\subsection{NGTS}
\begin{table}
	\centering
	\caption{Stellar Properties and colour magnitudes for \Nstar\ obtained from Gaia DR2 \citep{GAIA2018}, NGTS \citep{Wheatley2018} TIC v8 \citep{Stassun2019} and 2MASS
	\citep{Skrutskie2006}.}
	\label{tab:stellar}
	\begin{tabular}{ccc} 
	Property	&	Value		&Source\\
	\hline
	Gaia I.D.		&	DR2 5678383069566263552	& Gaia \\
    TIC I.D.		&	176772671	& TIC v8 \\
    R.A. (J2000)		&	09:30:16.0 				&	NGTS	\\
	Dec	 (J2000)		&	-18:50:33.62	&		  NGTS \\
    $\mu_{\alpha}$ (\masy)& $-30.528 \pm0.255$  & Gaia\\
    $\mu_{\delta}$ (\masy) & $18.0662 \pm0.234$ & Gaia\\
    Parallax (mas) & $4.392 \pm0.140$ & Gaia\\
    $G$			&$14.8357$	&Gaia\\
    NGTS		&$13.98$				& NGTS\\
    TESS		&$13.8995$				& 2MASS\\
    $V$ 		&$15.529$				& 2MASS\\
    $J$			&$12.701$		&2MASS	\\
   	$H$			&$12.06$		&2MASS	\\
	$K_{s}$			&$11.869$		&2MASS	\\
    
	\hline
	\end{tabular}
\end{table}
\Nstar\ was initially identified in photometry from the Next Generation Transit survey (hereafter NGTS; \citealt{Wheatley2018}). NGTS is a wide-field photometric survey consisting of an array of 12 fully automated 20~cm telescopes operating at ESO's Paranal observatory in Chile. The facility has been operational since early 2016, and is optimised for observations of K- and M-type stars, with sensitivity in the 520 to 890~nm wavelength range. NGTS has a wide field of view (instantaneously covering 96 sq deg) and delivers high cadence (every $\sim$ 13 seconds) photometry with high precision (1mmag per hour for an I=14 magnitude star).

The optimization of NGTS for precise photometry of late spectral-type stars has allowed it to make several discoveries of interesting M-dwarf systems. These include the discovery the most massive planet orbiting an M-dwarf \citep{Bayliss2018} as well as the shortest period brown dwarf around a main sequence star (in this case an early M-dwarf) \citep{Jackman2019}. NGTS has also discovered M-dwarfs in double-lined eclipsing binaries (\citealt{Casewell2018, Acton2020}) and low mass stars in long period orbits around stars of other spectral types (\citealt{Gill2019a, Lendl2019}). M-dwarf stars continue to be a key focus of the NGTS science program.

\Nstar\ was observed during the 2016 NGTS observing season. The field containing the system was observed for 156 nights between 2016 October 17\textsuperscript{th} and 2017 June 21\textsuperscript{st}, and in total we obtained 185,227 10\,s exposure science images. The magnitude of the system in various bandpasses, as well as positional information, is provided in Table \ref{tab:stellar}. To allow for the detection of the system, the light curve was first cleaned using an implementation of the SysRem algorithm \citep{2005MNRAS.356.1466T}. Periodic signals which do not show a typical transit shape (such as those caused by stellar variability) were then automatically detected and removed. After cleaning the eclipses were detected using \textsc{orion} (see \citealt{Wheatley2018} for more information), an implementation of the BLS algorithm \citep{Kovacs2016}.

\textsc{orion} also calculated some initial parameters for the system, identifying a 2.3\% depth eclipse with a period of 1.33 days, which allowed the object to be identified as a candidate exoplanet. When phase-folded on the \textsc{orion} identified period, we saw no evidence for a secondary eclipse at phase 0.5, implying that the companion must have a surface brightness which is significantly less than the primary star. Due to the lack of secondary eclipse and a transit depth which was consistent with a planetary companion, the object was followed-up photometrically and spectroscopically, where it was determined that the eclipsing object was in fact a low mass stellar companion.

\begin{figure}
	\includegraphics[width=\columnwidth]{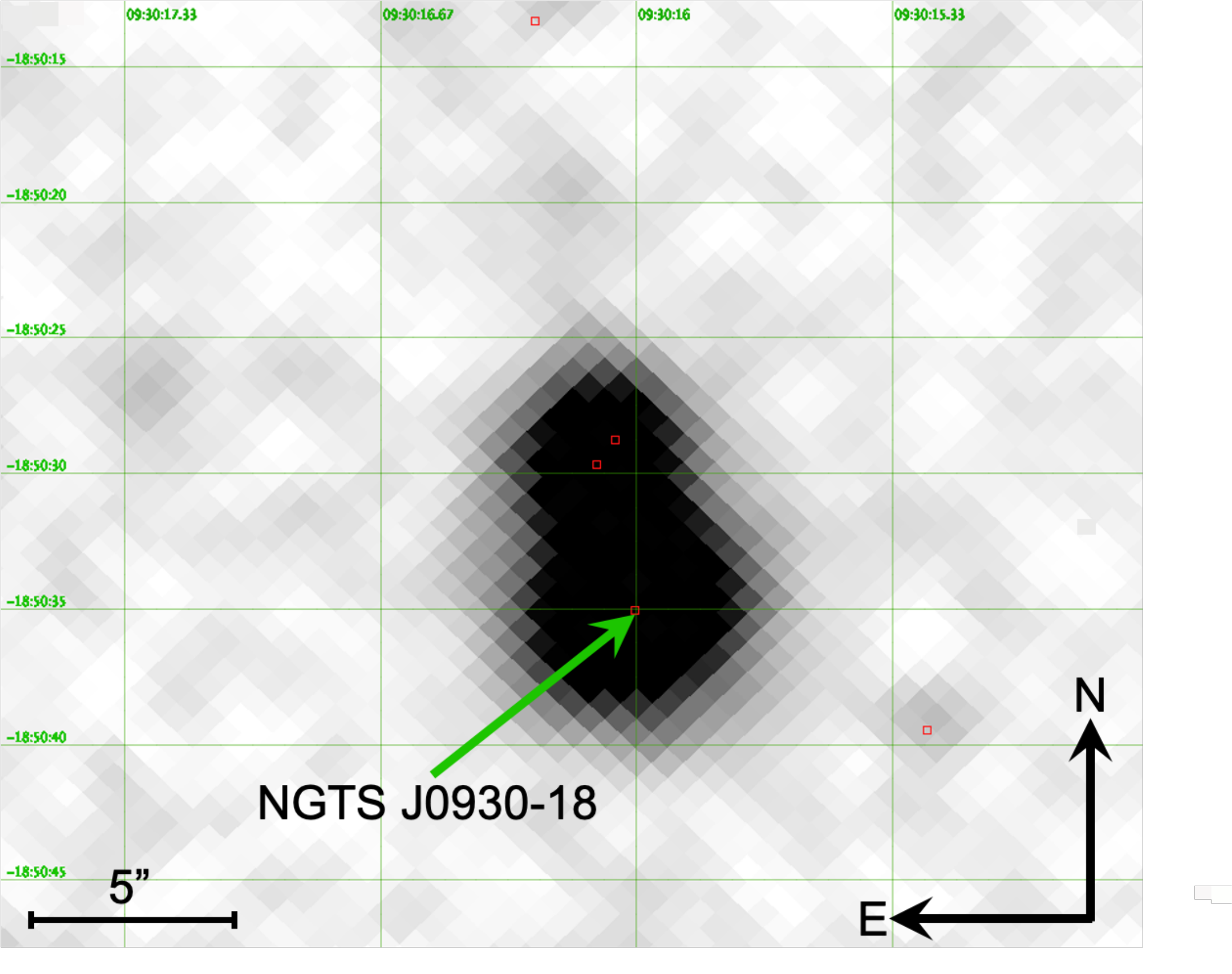}
    \caption{Digital Sky Survey (DSS) optical image of \Nstar. The red squares indicate the positions of identified Gaia DR2 sources. Four sources are present in the NGTS aperture which could contribute to the detection. \Nstar\, is the larger southern star, which is physically associated with the smaller northern star.}
    \label{fig:dss}
\end{figure}

Four Gaia DR2 \citep{GAIA2018} sources are present in the 15\arcsec radius photometric aperture used by the NGTS pipeline, consequently we could not be certain which star was the source of the eclipses identified by \textsc{orion} (see Figure~\ref{fig:dss}). Two of these stars are fainter than G=18 (Gaia band), and therefore contribute negligible flux and cannot be the source of the signal seen in the NGTS data. The remaining two objects have the same parallax and proper motion, and therefore are a physically related pair, with the eclipse event occurring on one of these. To identify the source of the eclipse we used the centroid vetting technique described in \citet{Gunther2017} which allows the measurement of extremely small shifts in the flux centroid during eclipse. Using this technique we identified a significant centroid shift indicating the southern star to be the eclipse source (Gaia ID - 5678383069566263552) - see Figure~\ref{fig:dss}.

\begin{figure}
	\includegraphics[width=\columnwidth]{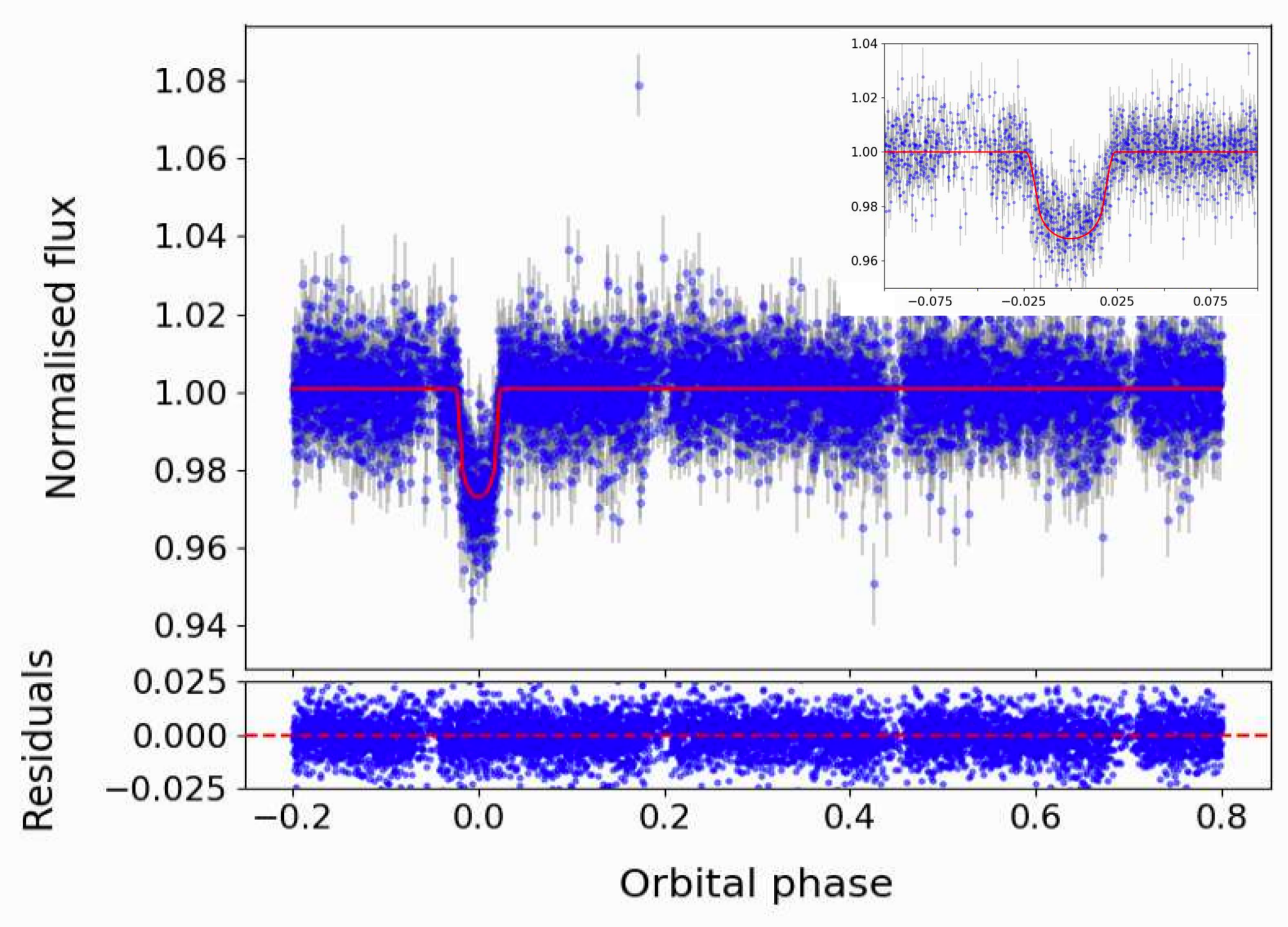}
    \caption{NGTS photometry of \Nstar\ folded on a period of \period\, days and binned to 5 minutes. The red line shows the model fit obtained from joint modelling of photometric and spectroscopic data.Note that there is no obvious secondary eclipse at phase 0.5. Inset --  zoomed in plot of the primary eclipse of the system.}
    \label{fig:ngts_lc}
\end{figure}

NGTS observations captured 14 full eclipses of the system in total, as well as a large number of partial eclipses at the start or end of observing nights. The NGTS discovery lightcurve is shown in Figure~\ref{fig:ngts_lc}.

\subsection{TESS}

\begin{figure}
	\includegraphics[width=\columnwidth]{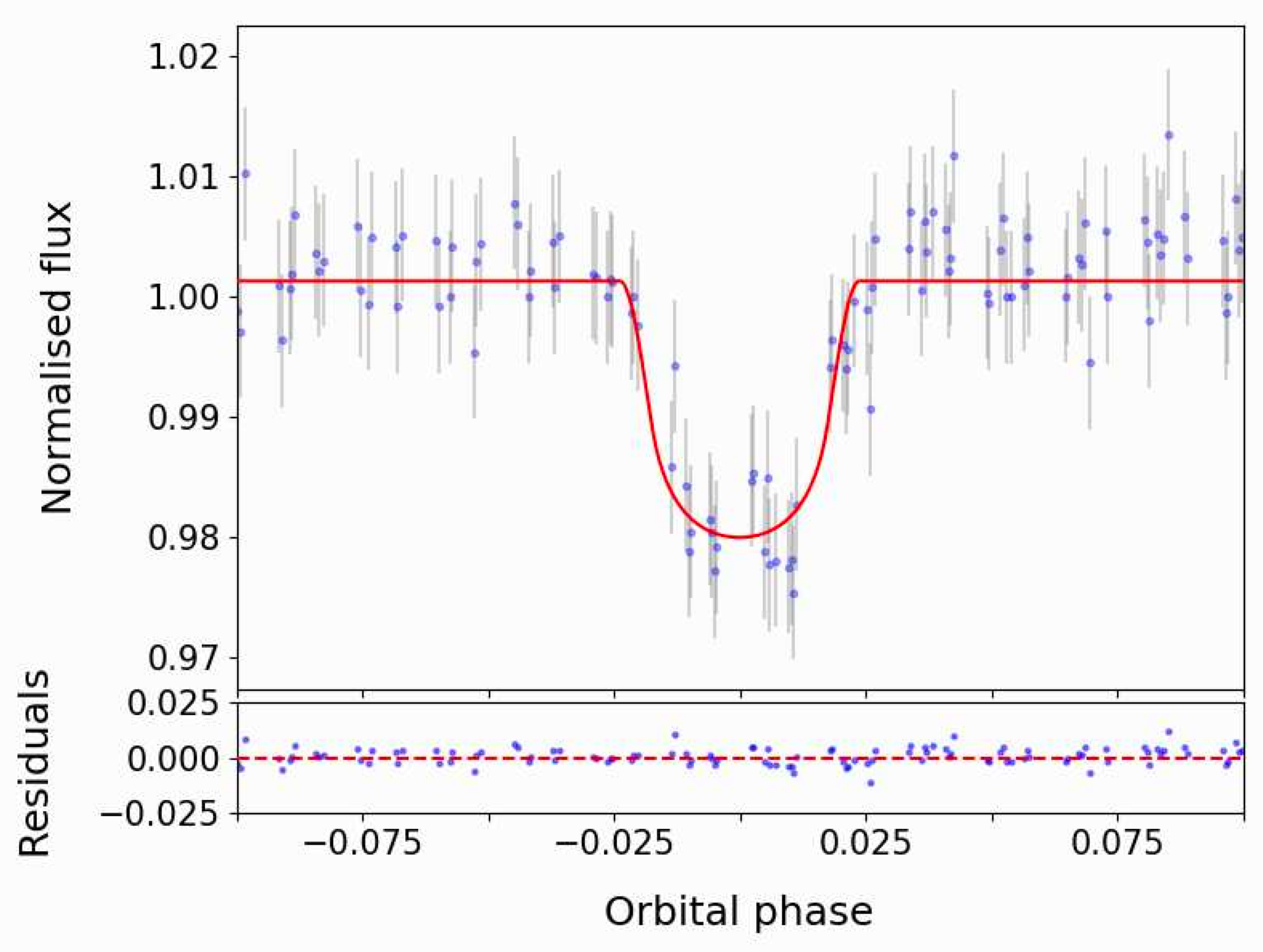}
    \caption{TESS 30-minute cadence photometry of the eclipse of \Nstar\ folded on the period of \period\, days. The red line shows the model fit obtained from joint modelling of photometric and spectroscopic data.}
    \label{fig:tess}
\end{figure}
\Nstar\ was observed in Sector 8 of the TESS mission (TIC-176772671, T=13.98), which occurred between 2019 February 2\textsuperscript{nd} and 2019 February 27\textsuperscript{th}. These observations consist of standard full frame images taken with a cadence of 30 minutes. We extracted the photometry for \Nstar\ from the full frame images of CCD 3 of camera 2. We used a custom aperture selected based on a flux threshold to minimise blending. However, due to the large TESS pixel size the blending is still greater than in the NGTS lightcurve. A floating median was applied to mask out systematic flux drops due to spacecraft effects.  Further details of this method can be found in \citealt{Gill2020}.

The signal is clearly detected in the TESS data, with a total of seven eclipses captured by the TESS observations. A BLS search of the TESS data alone identifies a similar periodicity to the NGTS data, further validating this detection. The TESS lightcurve is shown in Figure~\ref{fig:tess}. As for the NGTS lightcurve, we see no evidence of a secondary eclipse, or significant out of transit variation in the lightcurve.

\subsection{SAAO Photometry}
We obtained photometry of \Nstar\ using the SAAO 1-m telescope equipped with the SHOC instrument \citep{Coppejans2013} on 2018 December 20\textsuperscript{th} and 2019 January 29\textsuperscript{th} in I and g' bands. The aim of the observation was to confirm transit depth and width and to refine the ephemeris for the system. Additionally, by obtaining multi-colour photometry we can check for any wavelength dependent eclipse depth differences. The observations would also allow us to confirm which star is the source of the transit signal as they should be spatially resolved given the plate-scale of SHOC on the SAAO 1-m telescope (0.167\arcsec per pixel, binned by a factor 4).  The I band observations consisted of 360x30~sec exposures for a total observation time of 3 hours. The g' band observations were 770x10~sec exposures for a total observing time of 7700~seconds ($\equiv$2.13 hours). 

Standard bias and flat fielding corrections were applied to the data using the local SAAO SHOC pipeline, which is driven by {\sc python} scripts running {\sc iraf} tasks ({\sc pyfits} and {\sc pyraf}). Aperture photometry was performed using the {\sc Starlink} package {\sc autophotom}, which also measured and subtracted the sky background. The number of comparison stars and size of the aperture were chosen to minimise the RMS scatter outside of the eclipse. For both sets of observations we used a 4 pixel (2.67\arcsec) aperture with 2 stars used for comparison. 

\begin{figure}
	\includegraphics[width=\columnwidth]{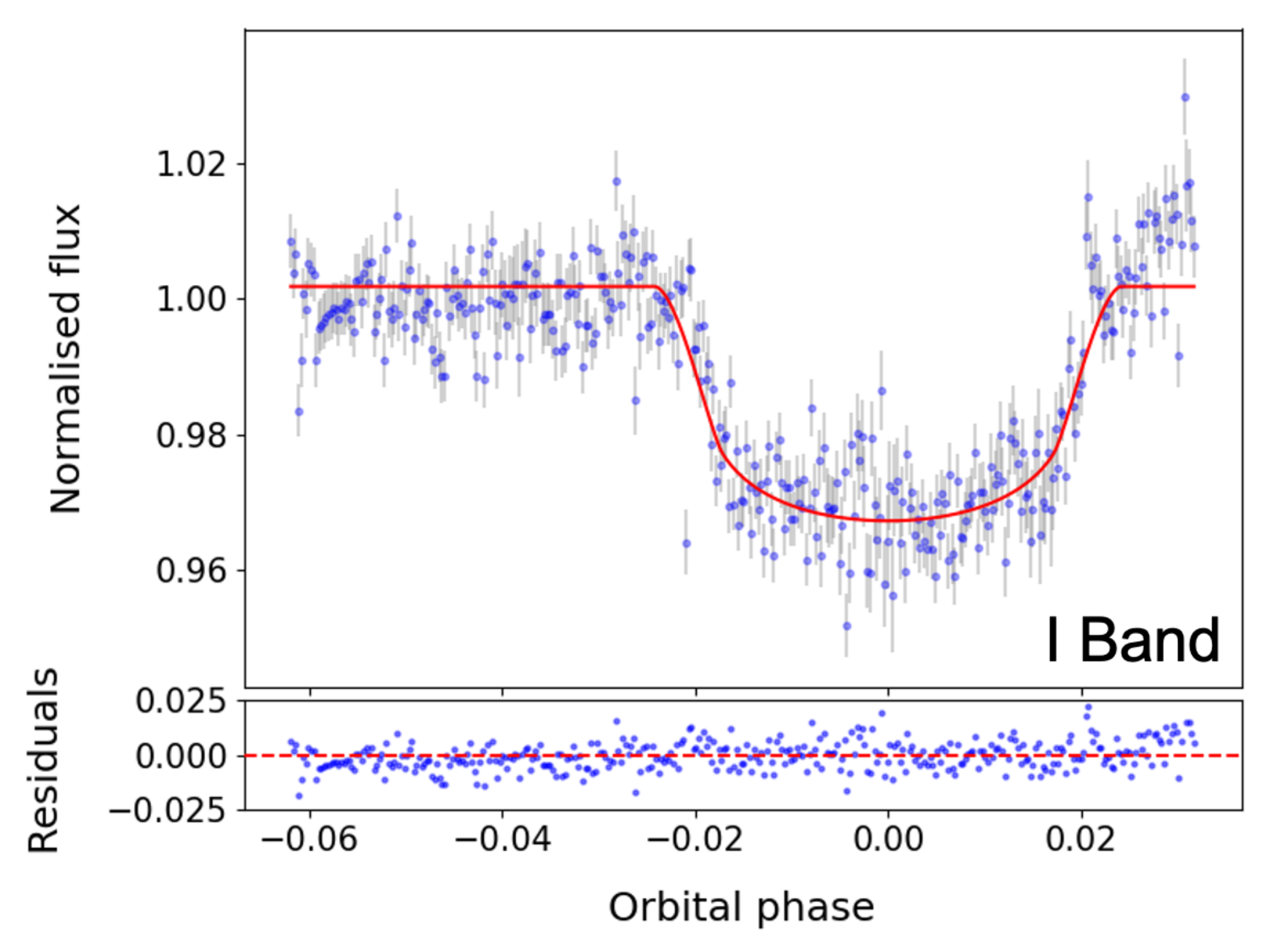}
    \caption{SAAO 1\,m/SHOC $I$-band photometry from the night 2018 December 20\textsuperscript{th} showing a primary eclipse of \Nstar\ plotted in phase. The red line shows the model fit obtained from joint modelling of photometric and spectroscopic data.}
    \label{fig:I_lc}
\end{figure}

\begin{figure}
	\includegraphics[width=\columnwidth]{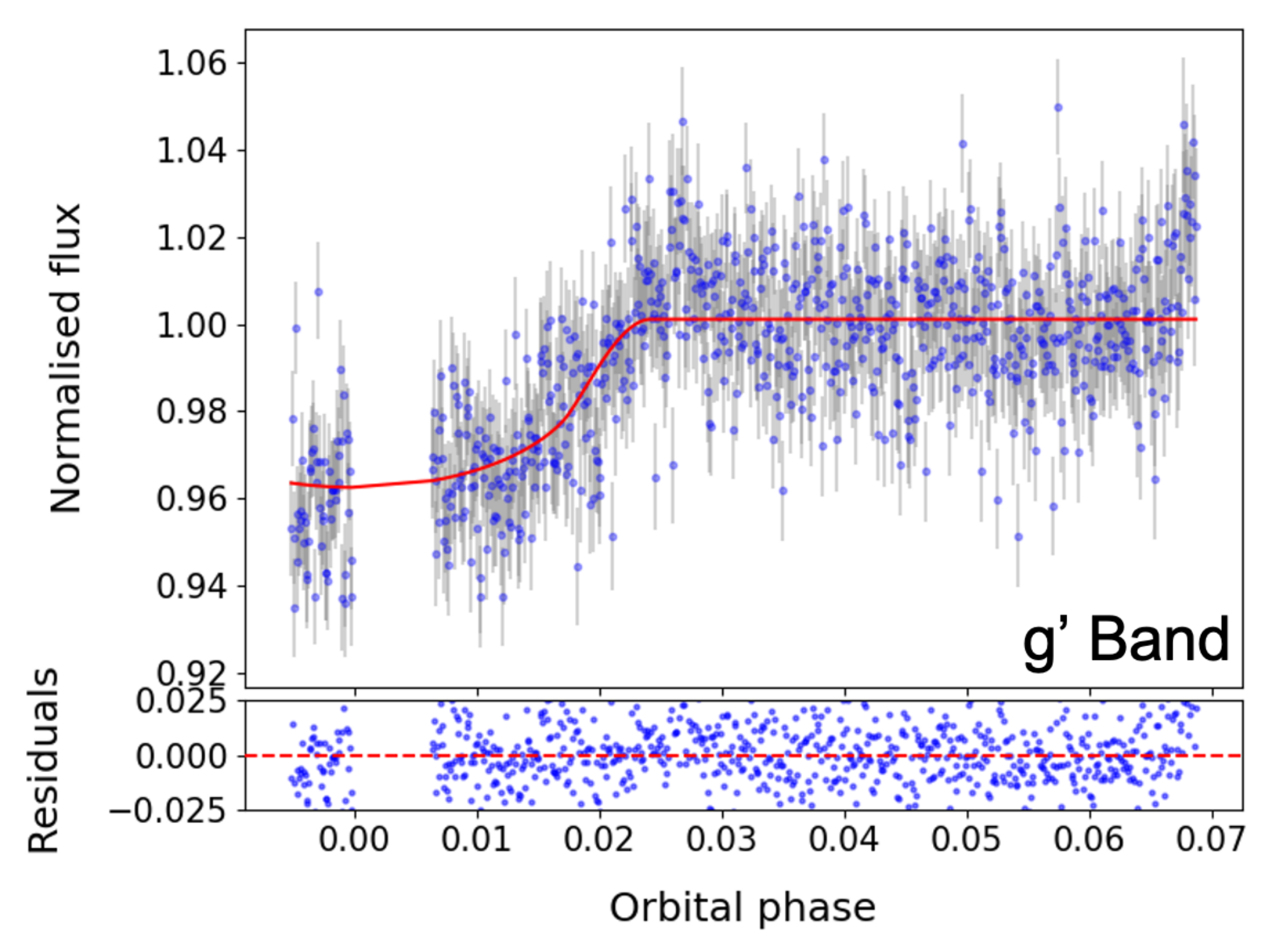}
    \caption{SAAO 1\,m/SHOC $g'$ filter photometry from the night 2019 January 29\textsuperscript{th}  of a primary eclipse of \Nstar\ plotted in phase. The red line shows the model fit obtained from joint modelling of photometric and spectroscopic data. The data gap starting near phase 0.0 is due to passing clouds during the observations.}
    \label{fig:g_lc}
\end{figure}

Both of these observations clearly detect the eclipse. The I band observations detect a full eclipse of the system, whereas the g' band observations capture the flux during eclipse as well as the egress. Additionally, the two brighter Gaia sources were able to be resolved. From this we were able to confirm that the eclipse occurred on the southern star, the brighter of the two possible eclipse sources, consistent with the result from centroid analysis. The lightcurves for each filter are shown in Figure \ref{fig:I_lc} and Figure \ref{fig:g_lc}. We note that there is not a significant difference in the eclipse depth between the two bands.

\subsection{Radial Velocity Measurements}
To determine the mass of the eclipsing object, \Nstar\ was observed with the HARPS spectrograph on the ESO 3.6m telescope \citep{Mayor2003} under programme 0103.C-0719 (PI. Bouchy). Due to the optical faintness of the object (V = 15.30) we used the high efficiency (EGGS) mode, which uses a larger fibre in order to improve S/N at the expense of a modest reduction in spectral resolution. A total of six observations were taken between 2019 April 11\textsuperscript{th} and 2019 June 8\textsuperscript{th}, with all exposure times being 2700 seconds.

\begin{figure}
	\includegraphics[width=\columnwidth]{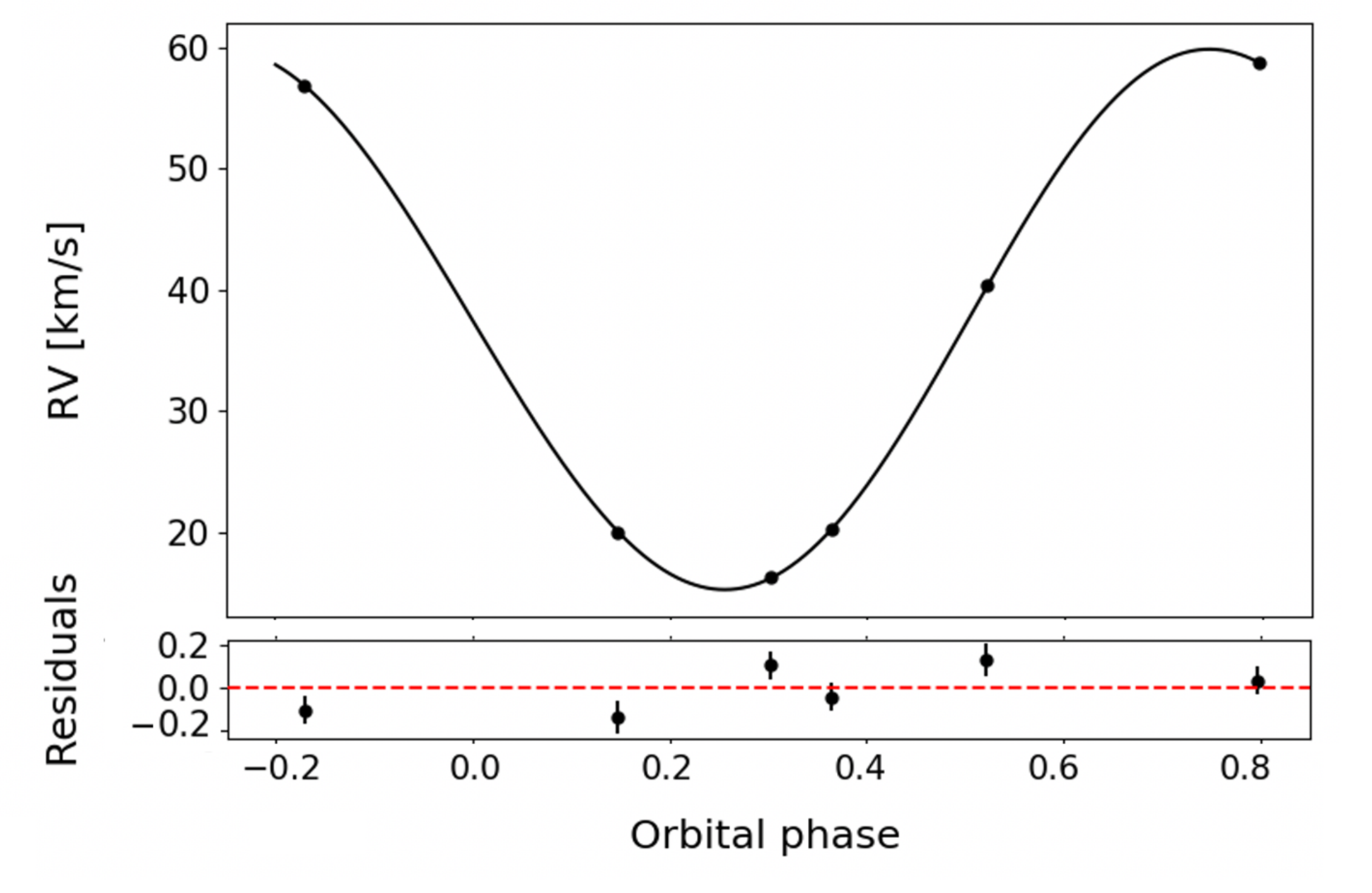}
    \caption{Phase folded radial velocity curve for \Nstar\, with black circles showing the six radial velocity measurements taken by HARPS between 2019 April 11\textsuperscript{th} and 2019 June 8\textsuperscript{th}. Radial velocities were phase folded on a period of \period\, days.  The solid black line shows the model fit to the radial velocities obtained from global modelling of the system. Fit residuals (data$-$model) are shown in the lower panel.}
    \label{fig:rvs}
\end{figure}

Observations were reduced using the standard HARPS data reduction pipeline.  The spectra were cross-correlated with a template G2 stellar mask and the cross correlation function (CCF) derived to determine the radial velocity of the star for each observation epoch. We used a G-type mask rather than an M-type as the star is rapidly rotating, and M-type masks struggle to deal with this. A single peak with large (K $\sim$ 22 km/s scale) velocity shift was detected, consistent with a low mass stellar companion. The radial velocities show a clear periodicity in phase with the period determined from photometric observations. The phase folded radial velocity curve is shown in Figure~\ref{fig:rvs}. We checked for correlation between bisector span and radial velocity and found no evidence for such correlation.  The full radial velocity measurements are given in Table~\ref{tab:RV_summary}.

\begin{table*}
	\centering
	\caption{Radial Velocities for \Nstar}
	\label{tab:RV_summary}
	\begin{tabular}{cccccc} 
    \hline
BJD$_\mathrm{TDB}$			&	RV		&RV error &	FWHM& 	Contrast& \\
(-2,450,000)	& (km/s)& (km/s)& (km/s)&(\%) & \\
		\hline
8584.6529	& 20.269  &	0.064&	33.650&	12.602\\
8585.6940	& 20.025  &	0.077&	33.957&  12.045\\
8586.6039  & 56.840  &	0.065&	32.117&	12.036\\
8638.5344  &58.792  &	0.062&	32.464&	11.472\\
8640.5395   &16.267   & 0.065&  32.790&  12.812\\
8643.4978   &40.395   &   0.074&  33.058&  12.180\\ \hline

	\end{tabular}
\end{table*}

\section{Results}

\subsection{Primary Star Parameters}

\subsubsection{Spectral Typing}\label{SpecTyping}
In order to determine the spectral type of the primary star, \Nstar\,A, we performed a template matching procedure. The HARPS spectra were wavelength shifted and then co-added to create a single higher signal-to-noise spectrum for analysis (SNR of 17). 
The spectral type was determined by comparing this spectrum with templates derived from Sloan Digital Sky Survey (SDSS) spectra covering a wide range of spectral types using the \textsc{pyhammer} code \citep{Kesseli2017} a python implementation of the \textsc{hammer} spectral classification routine \citep{Covey2007}.  \textsc{pyhammer} computes a chi-squared value that compares 34 spectral indices for each template, which are weighted relative to uncertainties in the individual observed spectra, to the corresponding values for the input spectrum. The spectral type which produces the minimum chi-squared value is taken as the assumed spectral type of the input star. The best matching template to our combined HARPS spectrum is that of an M0V star with [Fe/H] $\sim$ 0. We note however that the metallicity is not strongly constrained by these fits, with higher and lower metallicity templates also showing reasonable agreement with our spectrum.

\begin{table}
	\centering
	\caption{\Nstar\,A Stellar parameters derived from {\sc specmatch-emp} and {\sc ariadne}. For the parameters derived by {\sc ariadne}, the first error is the statistical error whilst the second is a systematic error calculated from the maximum difference between the average value and the values produced by the individual theoretical models}
	\label{tab:stellar_params}
	\renewcommand{\arraystretch}{1.5}
\begin{tabular}{ccc}
\hline
     Parameter &              {\sc specmatch-emp} &   {\sc ariadne}            \\ \hline
     Teff (K) &  4069 $\pm$  70 &  3982$^{+15}_{-16}$     $\pm$64\\
     Logg ($cm s^{-2}$)&  4.67 $\pm$0.12 &  4.687 $^{+0.055}_{-0.055}$  $\pm$0.045  \\
     Radius ($R_{\odot}$)&  0.62 $\pm$0.10 &   0.584 $^{+0.0094}_{-0.0103}$ $\pm$0.020\\
     $\left[Fe/H\right]$ &  -0.01 $\pm$0.09 & -0.012 $^{+0.041}_{-0.046}$ $\pm$0.085\\
     Mass ($M_{\odot}$)&  0.62 $\pm$0.08 &  0.580   $^{+0.0092}_{-0.0063}$ $\pm$0.017\\
     Age~(Gyr) &  9.89 $\pm$  0.17 &  9.20$^{+2.20}_{-5.31}$ $\pm$3.63\\
     Distance~(pc) & ---- &  223.5$^{+3.5}_{-3.0}$ $\pm$4.5 \\
     \hline
\end{tabular}
	
\end{table}

\subsubsection{Spectral Analysis}\label{specmatch}
An initial estimate of the stellar parameters for \Nstar\,A was obtained using \textsc{specmatch-emp} \citep{Yee2017}. These parameters would be used as priors when fitting the spectral energy distribution of the star to improve the quality of the result. \textsc{specmatch-emp} characterises stars based on their optical spectra, making use of a substantial library of high resolution (R$\sim$55,000), high S/N (>100) spectra obtained using Keck/HIRES. These high quality template spectra are used to classify an input spectrum (in this case the combined HARPS spectrum of \Nstar\,A).

\textsc{specmatch-emp} effectively performs a two step process. First, the input spectrum is shifted so that it is on the same wavelength scale as the library (template) spectra. This is achieved by performing a cross correlation between the input spectrum and several reference spectra in turn for a predetermined wavelength region. The reference spectrum which gives the largest cross correlation peak is then used to shift the entire spectrum. 

Once the HARPS spectrum has been shifted to the appropriate wavelength range, the matching procedure is applied. The input spectrum is compared with every other star in the library (for a given wavelength range, e.g., Mg~b triplet). $V\sin i$ is allowed to float, and a spline is fit to the continuum. The best matching stars to the input spectrum are identified using a chi-squared analysis. Linear combinations of the best matching spectra are then used to obtain an even better match than the individual best matching spectra alone. A weighted average of the library parameters is then taken and used to determine the properties of the target star. For \Nstar\,A these are given in Table~\ref{tab:stellar_params}.

Additionally we measured the projected stellar rotation velocity (v sin(i)) by fitting synthesised spectra using iSpec \citep{Blanco2014}. We fit only for v sin(i), fixing the other values to those obtained from \textsc{specmatch-emp}. This provides a value for the projected stellar rotation velocity of 30.17 $km s^{-1}$, showing that the star is rapidly rotating. Given the inclination determined in \ref{Fitting}, this is consistent with the star being in a state of spin orbit synchronisation.

\subsubsection{SED Fitting}\label{ariadne}
We fit the spectral energy distribution (hereafter, SED) of \Nstar\,A using \textsc{ariadne} \citep{Vines2020}. \textsc{ariadne} is a python tool which fits catalogue photometry to various atmospheric model grids. We fitted model grids from \texttt{Phoenix v2} \citep{Husser2013}, \texttt{BT-Settl}, \texttt{BT-Cond}, \texttt{BT-NextGen} \citep{Allard2012, Hauschildt99}, \cite{Castelli2004}, and \cite{Kurucz1993}. These were then convolved with various filter response functions. 

Each model SED is created by interpolating the model grids in \teff\--\logg\--\feh\, space. We also used distance, radius and extinction in the $V$ band as model parameters. Additionally we include an excess noise term for each set of photometry to account for any underestimation of uncertainties. We applied priors for \teff\,, \logg\, and \feh\, derived from the {\sc specmatch} analysis of the stacked HARPS spectrum (see section \ref{specmatch}). Priors for radius and distance were taken from Gaia DR2 and  $A_{V}$ was limited to the maximum line of sight value from the Schlegel, Finkbeiner \& Davis (SFD) galactic dust map (\citealt{Schlegel1998, Schlafly2011}). Excess noise parameters were normally distributed around zero with a variance equal to five times the size of the reported uncertainty. 

\begin{figure}
	\includegraphics[width=\columnwidth]{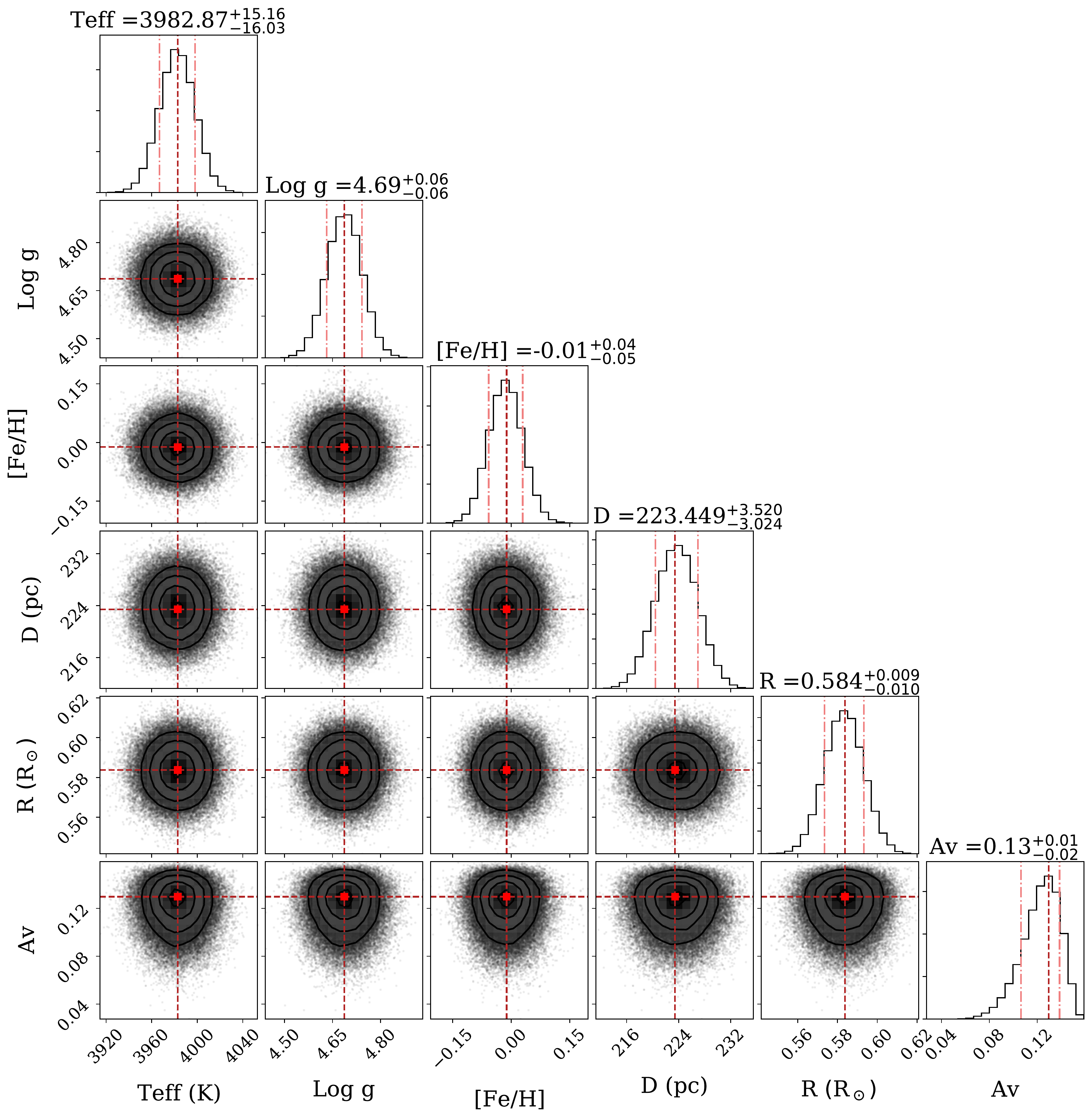}
    \caption{Cornerplot showing Key Primary Star parameters derived using \textsc{ariadne} in section \ref{ariadne}}
    \label{fig:ariadnecorner}
\end{figure}

Parameter estimation was performed using \texttt{dynesty}'s nested sampler for parameter estimation and calculating the bayesian evidence for each model \citep{Speagle2019}. \textsc{ariadne} then calculates the weighted average of each parameter using the relative probabilities of each of the fitted models. A mass estimate is then computed using MIST isochrones \citep{Choi2016}. A detailed explanation of \textsc{ariadne} is given in \citet{Vines2020}. The parameters for \Nstar\,A derived by \textsc{ariadne} are given in Table \ref{tab:stellar_params} and a corner plot showing the posterior distribution of the key parameters is shown in Figure \ref{fig:ariadnecorner}.

We note that the method employed by \textsc{ariadne} leads to results with a remarkable degree of precision. This is a result of the mathematical treatment of the posterior parameter distribution. The distribution of each model is averaged, using the relative probabilities of each model as weights. This is determined using the following equation

\begin{equation}
    P(\theta_i)=\sum_{n=1}^{N} P(\theta \mid X, M_n) P(M_n \mid X)
\end{equation}

Where $\theta_i$ is the parameter to be averaged, $P(\theta \mid X, M_n)$ is the posterior distribution derived using Bayes Theorem and $P(M_n \mid X)$ is the Bayesian evidence of the individual model. This probability is used as a weight when averaging over the full set of models.

\begin{figure}
	\includegraphics[width=\columnwidth]{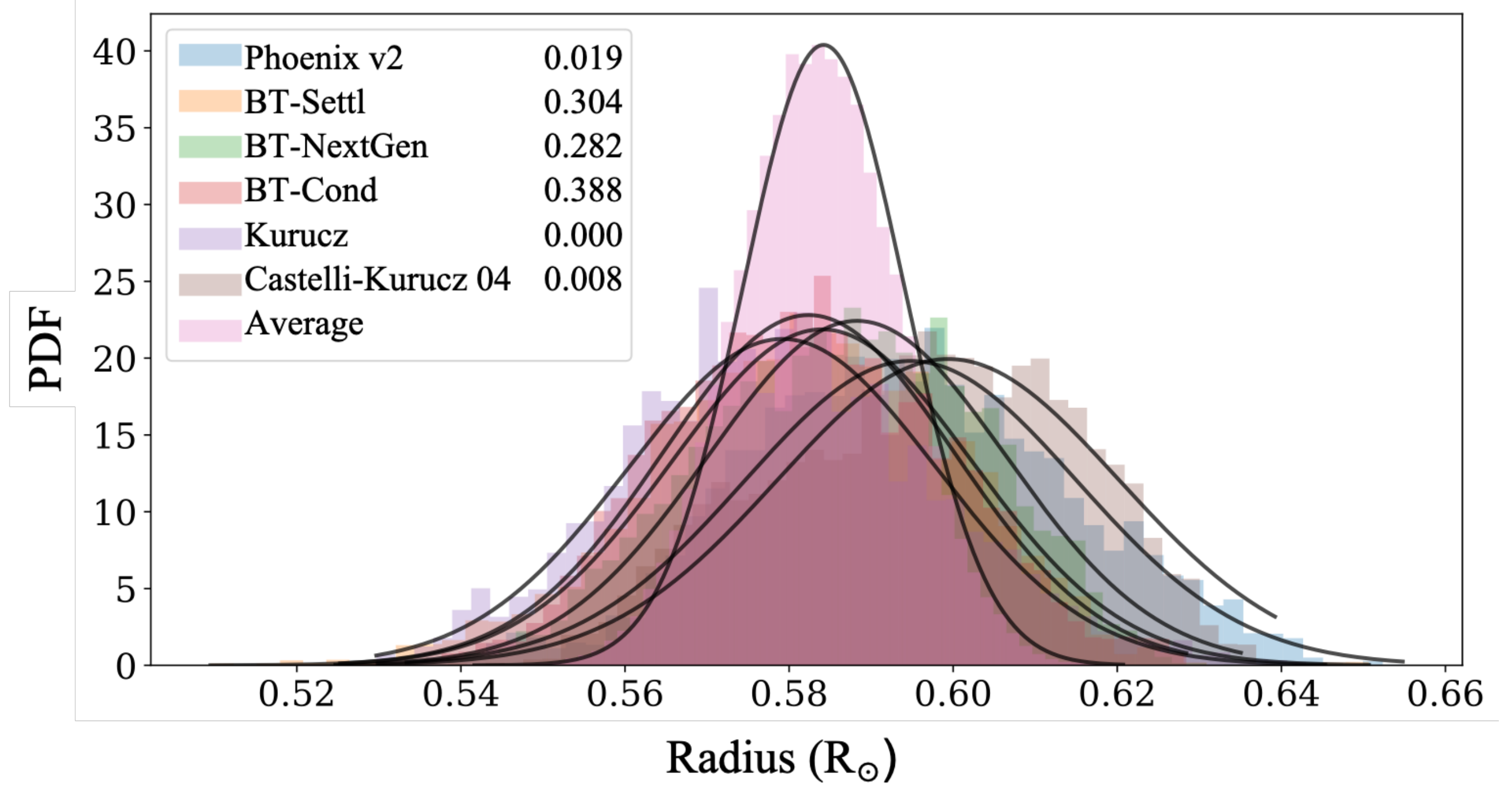}
    \caption{Histogram showing the probability distribution for the radius of \Nstar\,A for each atmospheric model. The model probabilities are then used as weights to compute an overall average estimation, shown in pink, which has the best model value and uncertainty for the parameter.}
    \label{fig:ariadnehistogram}
\end{figure}

This results in higher precision than would be obtained by the use of any one model alone. The increased precision obtained by averaging over posterior distributions can be seen in Figure \ref{fig:ariadnehistogram}, which shows the probability density function for the derived stellar radius. Preliminary testing has shown that \textsc{ariadne} obtains great accuracy compared to radii derived by interferometry. This is particularly important due to the direct dependence of the mass and radius of the companion on that of the host. 

To account for the effect of underestimating uncertainties when averaging over models, we also calculate a systematic error for the parameters derived by \textsc{ariadne}. As in \cite{Southworth2015}, we determine this by fitting the SED with each model individually, and taking the largest difference between these individual values and the average value as the systematic error.

The values obtained with \textsc{ariadne} are more precise, but still consistent with those obtained by \textsc{specmatch-emp}. These parameters are roughly consistent with an early M-dwarf star of spectral type M0V, as determined in \S\ref{SpecTyping}. We adopt the parameters derived by \textsc{ariadne} to derive the orbital solution as detailed in \S\ref{Fitting}.

\subsection{Global Modelling}\label{Fitting}

\begin{table*}
\renewcommand{\arraystretch}{1.35}
	\centering
	\caption{Fitted and Derived parameters for \Nstar. For parameters derived using values from \textsc{ariadne}, we also report a systematic uncertainty as described in Section \ref{ariadne}}
	\label{tab:fit_params}
	\begin{tabular}{ccccc} 
    \hline
Quantity & Description  & unit & Value & Error\\
\hline
Fitted parameters \\ 
{$\frac{R_{\rm pri}}{a}$} & radius ratio of primary to semi-major axis & none & $0.1243$ & $^{+0.0023}_{-0.0016}$\\

$k$ & radius ratio of stars, $R_{\rm sec}/R_{\rm pri}$ & none & $0.1814$ & $^{+0.0019}_{-0.0020}$\\

$b$ & impact parameter, $a \cos{(i)}/R_{\rm pri}$ & none & $0.013$ & $^{+0.095}_{-0.011}$\\



$P$ & orbital period & days & $1.33264614$ & $^{+0.00000126}_{-0.00000090}$ \\

$T_{c}$ & epoch of primary eclipse centre & BJD & $2457679.29957$ & $^{+0.00053}_{-0.00057}$ \\

{$\sigma_{\rm NGTS}$} & systematic error in NGTS light curve & norm. flux & $0.00593$ & $^{+0.00010}_{-0.00012}$\\

{$\sigma_{I}$} & systematic error in I light curve & norm. flux & $0.00454$ & $^{+0.00027}_{-0.00040}$ \\

{$\sigma_{g'}$} & systematic error in g' light curve & norm. flux & $0.00838$ & $^{+0.00062}_{-0.00063}$ \\

{$\sigma_{\rm TESS}$} & systematic error in TESS light curve & norm. flux & $0.000045$ & $^{+0.000330}_{-0.000051}$ \\

{$\beta_{\rm NGTS}$} & normalised flux scale factor in NGTS data & none & $1.000908$ & $^{+0.000099}_{-0.000016}$ \\

{$\beta_{I}$} & normalised flux scale factor in I data & none & $1.00086$ & $^{+0.00036}_{-0.00044}$  \\

{$\beta_{g'}$} & normalised flux scale factor in g' data & none & $1.00049$ & $^{+0.00049}_{-0.00061}$  \\

{$\beta_{\rm TESS}$} & normalised flux scale factor in TESS data & none & $1.00155$ & $^{+0.00030}_{-0.00019}$  \\

$u_{\rm{NGTS}}$ & linear LDC in NGTS band & none & $0.499$ & $^{+0.035}_{-0.026}$ \\

$u'_{\rm{NGTS}}$ & quadratic LDC in NGTS band & none & $0.223$ & $^{+0.039}_{-0.064}$ \\

$u_{\rm{I}}$ & linear LDC in I band & none & $0.383$ & $^{+0.038}_{-0.032}$ \\

$u'_{\rm{I}}$ & quadratic LDC in I band & none & $0.434$ & $^{+0.056}_{-0.071}$ \\

$u_{\rm{g'}}$ & linear LDC in g' band & none & $0.500$ & $^{+0.029}_{-0.036}$ \\

$u'_{\rm{g'}}$ & quadratic LDC in g' band & none & $0.367$ & $^{+0.066}_{-0.043}$ \\

$u_{\rm{TESS}}$ & linear LDC in TESS band & none & $0.487$ & $^{+0.030}_{-0.030}$ \\

$u'_{\rm{TESS}}$ & quadratic LDC in TESS band & none & $0.175$ & $^{+0.032}_{-0.025}$ \\

{$K$} & radial velocity semi-amplitude & km/s & $21.975$ & $^{+0.404}_{-0.099}$  \\

{$\Gamma_{\rm pri}$} & systemic velocity & km/s & $37.2464$ & $^{+0.0070}_{-0.2443}$ \\

{$\sigma_{\rm RV}$} & jitter in RV data & km/s & $0.287$ & $^{+0.075}_{-0.260}$  \\


Derived parameters \\

{$R_{\rm sec}$} & radius of secondary & $R_\odot$ & $0.1059$ & $^{+0.0023}_{-0.0021}$ ($\pm$ 0.0040)\\

{$R_{\rm sec}$} & radius of secondary & $R_{J}$ & $1.052$ & $^{+0.023}_{-0.021}$  ($\pm$ 0.040) \\

$m_{\rm sec}$ & mass of secondary & $M_\odot$ & $0.0818$ & $^{+0.0040}_{-0.0015}$ ($\pm$ 0.0052)\\

$m_{\rm sec}$ & mass of secondary &$M_{J}$ & $85.7$ & $^{+4.2}_{-1.5}$ ($\pm$ 5.4)\\

{$q$} & stellar mass ratio & none & $0.1407$ & $^{+0.0065}_{-0.0017}$ ($\pm$ 0.0085)\\

{$a$} & semi-major axis of system & $AU$ & $0.02195$ & $^{+0.00040}_{-0.00064}$ ($\pm$ 0.00088)\\

$i$ & orbital inclination & $deg$ & $89.914$ & $^{+0.085}_{-0.671}$ \\


$T_{\rm 14-pri}$ & primary eclipse duration & hours & $1.510$ & $^{+0.021}_{-0.026}$ \\

	\hline
    \end{tabular}
\end{table*}

To determine the mass and radius of \Nstar\,B, as well as other system parameters, we performed global fitting of both the photometric data (NGTS, TESS and SAAO) and the HARPS radial velocities. This was performed using the eclipsing binary light curve simulation code {\textsc{ellc}} \citep{Maxted2016} in combination with the Markov Chain Monte Carlo (MCMC) sampler {\textsc{emcee}} \citep{Foreman-Mackey2013}. Before performing this, we normalised the raw lightcurves by their median out of eclipse flux, and binned the NGTS data to five minutes to reduce computational time.

The walkers were initialised in a region of parameter space which provided a good initial fit. Each walker was then given a starting position selected from a normal distribution centred on these values. We used the values derived by \textsc{orion} to obtain initial values for both the transit epoch and the orbital period. Initial values for the primary star radius, stellar radius ratio, impact parameter, light ratio and radial velocity components were determined by first running the MCMC for a small number of steps to find values that gave a reasonable initial fit.  We also incorporated a radial velocity jitter term added in quadrature in our modelling to account for stellar noise, as well as normalisation scaling parameters and systematic errors for each of the four lightcurves. We additionally fit for a third light parameter to account for dilution in the NGTS and TESS lightcurves. We fixed this parameter to zero for the SAAO lightcurves due to the fact that the neighbouring star was not present in the aperture used for reduction.

Limb darkening parameters were obtained using the \textsc{ldtk} software \citep{Parvianen2015}. A quadratic limb darkening law was used with stellar properties, e.g., \teff\, and \logg\, taken from the results given by {\sc ariadne}. Limb darkening coefficients and uncertainties were calculated directly with \textsc{ldtk}, for each photometric filter used, and placed as priors for the fitting process.

An initial fit to the data resulted in an orbit with an eccentricity of \ecc. \cite{Lucy1971} showed that many systems with low levels of eccentricity are actually circular orbits, where the addition of eccentricity has improved the fit due to the introduction of additional free parameters. Using the methods in \cite{Lucy1971}, we determine that there is an 83\% probability that the measured eccentricity would be detected in a system with a true eccentricity of zero. Therefore for the final system parameters we force a circular fit.

We used 200 walkers and 50000 steps to model the lightcurve using the \textsc{emcee} sampler. Each walker used initial parameters that were randomized in a Gaussian ball around values previously determined to give a good initial fit. We note that alterations to our defined priors did not preclude the ability of the model to obtain a good fit. 10000 steps were discarded as burn-in and not used when analysing the results of the modelling. The modal values of the posterior distributions were adopted as the most probable parameters, with the 68.3 percent (1$\sigma$) highest probability density interval as the error estimates.
 
 This global modeling resulted in mass and radius for \Nstar\,B of \NstarMass $M_{\odot}$ (\NstarMassJup $M_{J}$) and \NstarRad $R_{\odot}$ (\NstarRadJup $R_{J}$).  The binary has a notably low mass ratio of just \massratio. The full list of best fitting parameters derived from the posterior distributions are given in Table \ref{tab:fit_params}.

\section{Discussion}
\begin{figure*}
	\includegraphics[width=\textwidth]{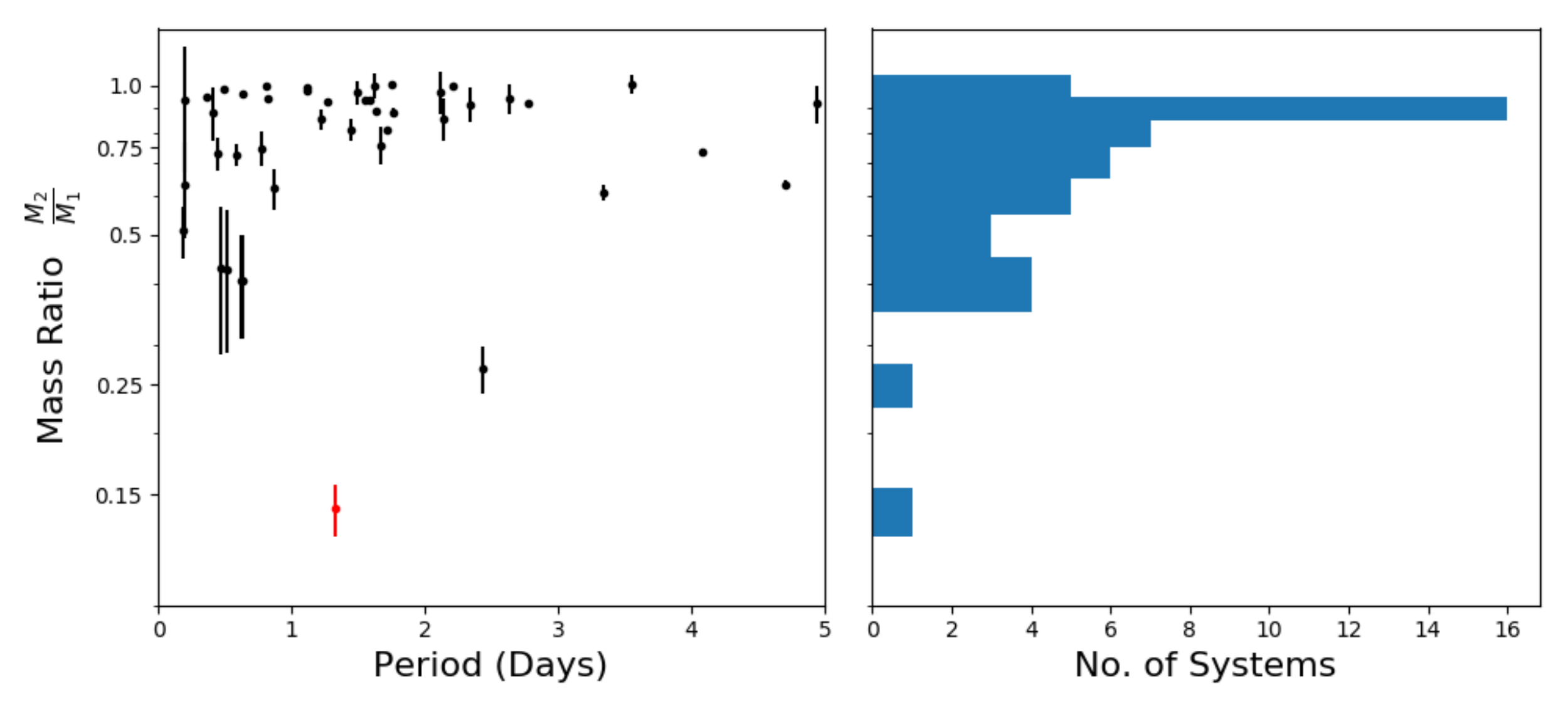}
    \caption{Left -- Mass ratio as a function of period for M-dwarf binaries given by \citealt{Parsons2018} and \citealt{Nefs2013}. \Nstar\ is indicated in Red. Right -- Histogram showing the distribution of M-dwarf binary mass ratios. \Nstar\ is a clear outlier from the general population.}
    \label{fig:massratios}
\end{figure*}
\subsection{Unequal mass M-dwarf binaries}

The results from SED fitting (\S\ref{ariadne}) indicate that \Nstar\,A has a mass of \PristarMass$M_{\odot}$ and a radius of \PristarRad $R_{\odot}$ i.e. an early M-dwarf \citep{Kaltenegger2009}.  Thus the system \Nstar\ is an unequal mass eclipsing M-dwarf binary. This is an unusual configuration, as most known eclipsing M-dwarf systems (e.g \citealt{Parsons2018}) are in near equal mass binaries. With a mass ratio of \massratio\, \Nstar\ is highly unusual.  The mass ratio is more similar to the well-studied population of low mass M-dwarfs in orbit around F- G- K- stars (e.g., \citealt{Triaud2017}) or transiting brown dwarf systems around early/mid M-dwarfs (e.g., \citealt{Irwin2010,Johnson2011, Montet2015, Gillen2017, David2019}).

It is expected that short-period M-dwarf systems with unequal mass components will be very rare. During formation, mass is preferentially accreted onto the lower mass component of the binary \citep{Bate2002} which will over time drive the mass ratio towards unity and result in a system with roughly equal mass components. For short period, low mass systems this effect is expected to be even greater due to dynamical effects. An early post-collapse star forming cloud will fragment into a low number of multiple systems \citep{Goodwin2007}.  Dynamical decay and interactions within this collapsing cloud are biased against low mass components, which are typically ejected on a short timescale \citep{Anosova1986}. In interactions with higher mass stars this means the low mass star tends to be swapped with a higher mass replacement. Thus the mechanisms that produce short period binaries are biased to produce equal mass systems.

This theory is supported by observational evidence. \citealt{Delfosse2004} surveyed a large number of M-dwarf binaries in the Solar neighborhood. They found that for systems with a period less than 50~days, the distribution of the mass ratio peaked close to unity. The orbital period of \Nstar\, is significantly shorter than this at \period~days, meaning its formation and survival probes a sparse area of binary star parameter space. There are, however, examples of similar unequal mass binaries in the literature (e.g., \citealt{Nefs2013}).

In Figure~\ref{fig:massratios}\, we show the mass ratio as a function of period for short period (less than 5 days) systems with M-dwarf primaries taken from the samples in \citealt{Parsons2018} and \citealt{Nefs2013}. From Figure~\ref{fig:massratios}\, it is clear that \Nstar\, has the lowest mass ratio of any known M-dwarf binary system. This also demonstrates that the vast majority of short period systems of this type have a mass ratio close to 1, given the large cluster of systems in the top left of the plot. Such a short period system of two M-dwarfs with highly unequal masses is clearly unusual.

\subsection{M-dwarf Mass-Radius Relationship}

\begin{figure}
	\includegraphics[width=\columnwidth]{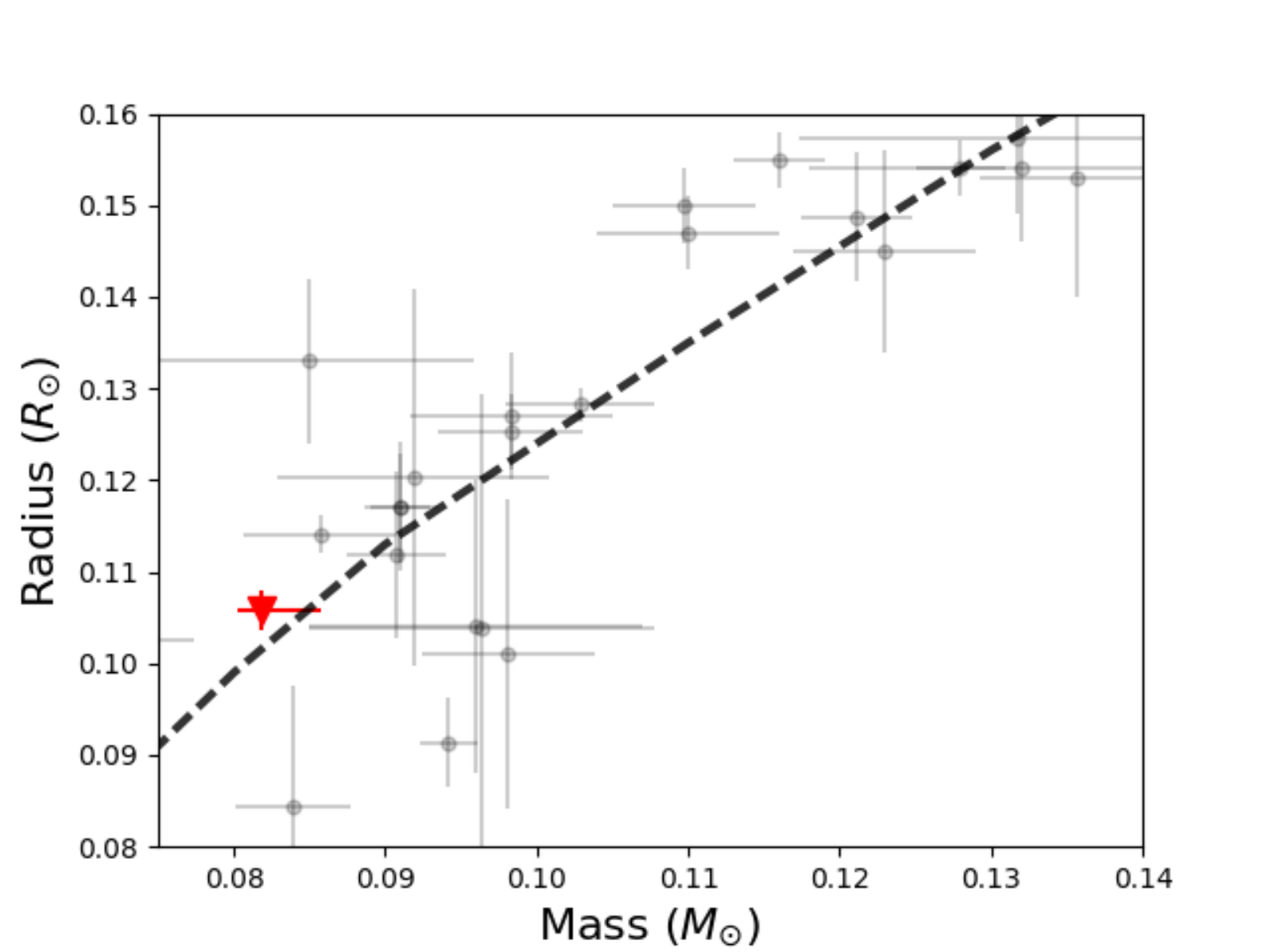}
    \caption{Comparison between \Nstar\,B and a model stellar isochrone from \citealt{Baraffe2015}. The 10-Gyr isochrone is indicated by the black dashed line, and \Nstar\,B by the red triangle. Similar M-dwarfs from \citealt{Parsons2018}, \citealt{Triaud2020b} and \citealt{Mireles2020} are shown in Black.
}
    \label{fig:massradius}
\end{figure}

The global modelling performed in section \ref{Fitting} indicates \Nstar\,B has a mass of \NstarMassJup M\textsubscript{J}. This places it just above the hydrogen burning mass limit of $\sim$ 70 M\textsubscript{J} which separates brown dwarfs and low mass stars \citep{Baraffe1998}. The mass and radius of the star are similar to that of TRAPPIST-1 \citep{Gillon2017}, demonstrating the importance of precise measurements in this parameter space to characterise current and future exoplanet discoveries.  


In Figure~\ref{fig:massradius}\, we compare the mass and radius of \Nstar\,B to a sample of low mass stars from \citealt{Parsons2018}, as well as a model 10-Gyr stellar isochrone from \citealt{Baraffe2015}.\@ Figure~\ref{fig:massradius}\, illustrates the low number of systems known in this parameter space. Most of the systems are from detections of low mass secondary stars in eclipsing binaries with higher mass primaries, the same configuration as \Nstar. We also note the well-known scatter in the mass--radius relationship for low mass stars is still present even at the lower end of the mass distribution. 


This shows that the radius of \Nstar\,B is very slightly inflated relative to models, but is still consistent to 1-sigma for its mass. Inflated radii of low mass stars can often be associated with short orbital periods \citep{Spada2013}. This is due to the fact that at fast rotation speeds magnetic activity is enhanced which can affect convective processes, causing an inflation in stellar radius (\citealt{Lopez2007, Chabrier2007}). However inflation of stellar radii has also been seen in longer period systems (e.g., \citealt{Acton2020}) indicating that this deviation is not consistent. Indeed, with a period of just \period\ days we may have expected this object to show some level of inflation, which is not clearly seen. We note that the HARPS spectrum of the star shows clear H$\alpha$ emission, a strong indicator that the star is indeed magnetically active.

We note that direct detection of the secondary eclipse of this system to refine these parameters in the future will likely be difficult. The non detection of a secondary eclipse in the NGTS data can be used to place an upper limit on the light ratio of the two stars. However given the out of eclipse scatter a secondary eclipse would still not be detected even if \Nstar\,B was more luminous than it actually is. For a 0.08$M_{\odot}$ dwarf with an age of 10 Gyr, \cite{Baraffe2015} give a temperature of 2345 K. From determining the surface brightness ratio for the system given the temperature of the primary determined in Section \ref{ariadne}, we find the secondary eclipse depth to be around 0.25\%, well within the scatter of both the TESS and NGTS lightcurves. An eclipse this shallow will be difficult to detect given the faintness of the system. Spectroscopic detection of a second set of lines associated with the secondary will also be difficult due to the rapid rotation of the star resulting in the blending of the spectral lines.

It is also important to note the precise mass and radius measurements obtained for \Nstar\,B when compared with similar objects in Figure \ref{fig:massradius}. Owing to the high precision photometry and radial velocities used to characterise the system we have derived the mass and radius of \Nstar\,B to a precision of  5$\%$ and 2$\%$ respectively. High precision measurements in this regime are vital for the empirical derivation of the mass--radius relationship stars at the lowest end of the mass radius distribution.

\section{Conclusions}
We have discovered the M-dwarf eclipsing binary system \Nstar, with the secondary component, \Nstar\,B, having a mass just above the hydrogen burning limit.  We were able to determine a very precise mass and radius for \Nstar\,B, and these parameters are of great scientific interest due to the prominence of very low mass stars in the search for temperate terrestrial exoplanets. Knowledge of the masses and radii of these stars are vital for characterisation of future exoplanet discoveries, and these measurements can only be obtained precisely through the characterisation of eclipsing binary stars similar to this system. \Nstar\,B provides a valuable data point in a sparsely populated region of parameter space and will be of importance for future work in empirically deriving the mass--radius relationship for the lowest mass stars.

\section*{Acknowledgements}
Based on data collected under the NGTS project at the ESO La Silla Paranal Observatory. The NGTS facility is operated by the consortium institutes with support from the UK Science and Technology Facilities Council (STFC) under projects ST/M001962/1 and ST/S002642/1. 

This paper includes data collected by the TESS mission. Funding for the TESS mission is provided by the NASA Explorer Program. 

This paper uses observations made at the South African Astronomical Observatory (SAAO). We thank Marissa Kotze (SAAO) for developing the SHOC camera data reduction pipeline. 

This study is based on observations collected at the European Southern Observatory under ESO programme 0103.C-0719. 

JA is supported by an STFC studentship. BTG, SG and PJW acknowledge support from STFC consolidated grants ST/L000733/1 and ST/P000495/1.MNG acknowledges support from MIT's Kavli Institute as a Juan Carlos Torres Fellow. JSJ acknowledges support by FONDECYT grant 1201371, and partial support from CONICYT project Basal AFB-170002.EG gratefully acknowledges support from the David and Claudia Harding Foundation in the form of a Winton Exoplanet Fellowship.

We thank the anonymous referee for their useful and constructive feedback to improve the paper.

\section*{Data Availability}
The data underlying this article will be shared on reasonable request to the corresponding author.



\bibliographystyle{mnras}
\bibliography{ref} 

\begin{thebibliography}{}
\makeatletter
\relax
\def\mn@urlcharsother{\let\do\@makeother \do\$\do\&\do\#\do\^\do\_\do\%\do\~}
\def\mn@doi{\begingroup\mn@urlcharsother \@ifnextchar [ {\mn@doi@}
  {\mn@doi@[]}}
\def\mn@doi@[#1]#2{\def\@tempa{#1}\ifx\@tempa\@empty \href
  {http://dx.doi.org/#2} {doi:#2}\else \href {http://dx.doi.org/#2} {#1}\fi
  \endgroup}
\def\mn@eprint#1#2{\mn@eprint@#1:#2::\@nil}
\def\mn@eprint@arXiv#1{\href {http://arxiv.org/abs/#1} {{\tt arXiv:#1}}}
\def\mn@eprint@dblp#1{\href {http://dblp.uni-trier.de/rec/bibtex/#1.xml}
  {dblp:#1}}
\def\mn@eprint@#1:#2:#3:#4\@nil{\def\@tempa {#1}\def\@tempb {#2}\def\@tempc
  {#3}\ifx \@tempc \@empty \let \@tempc \@tempb \let \@tempb \@tempa \fi \ifx
  \@tempb \@empty \def\@tempb {arXiv}\fi \@ifundefined
  {mn@eprint@\@tempb}{\@tempb:\@tempc}{\expandafter \expandafter \csname
  mn@eprint@\@tempb\endcsname \expandafter{\@tempc}}}

\bibitem[\protect\citeauthoryear{{Acton} et~al.,}{{Acton}
  et~al.}{2020}]{Acton2020}
{Acton} J.~S.,  et~al., 2020, \mn@doi [\mnras] {10.1093/mnras/staa928}, \href
  {https://ui.adsabs.harvard.edu/abs/2020MNRAS.494.3950A} {494, 3950}

\bibitem[\protect\citeauthoryear{Allard, Homeier  \& Freytag}{Allard
  et~al.}{2012}]{Allard2012}
Allard F.,  Homeier D.,   Freytag B.,  2012, \mn@doi [Philosophical
  Transactions of the Royal Society A: Mathematical, Physical and Engineering
  Sciences] {10.1098/rsta.2011.0269}, 370, 2765

\bibitem[\protect\citeauthoryear{{Anosova}}{{Anosova}}{1986}]{Anosova1986}
{Anosova} J.~P.,  1986, \mn@doi [\apss] {10.1007/BF00656037}, \href
  {https://ui.adsabs.harvard.edu/abs/1986Ap&SS.124..217A} {124, 217}

\bibitem[\protect\citeauthoryear{{Baraffe}, {Chabrier}, {Allard}  \&
  {Hauschildt}}{{Baraffe} et~al.}{1998}]{Baraffe1998}
{Baraffe} I.,  {Chabrier} G.,  {Allard} F.,   {Hauschildt} P.~H.,  1998, \aap,
  \href {https://ui.adsabs.harvard.edu/abs/1998A&A...337..403B} {337, 403}

\bibitem[\protect\citeauthoryear{{Baraffe}, {Homeier}, {Allard}  \&
  {Chabrier}}{{Baraffe} et~al.}{2015}]{Baraffe2015}
{Baraffe} I.,  {Homeier} D.,  {Allard} F.,   {Chabrier} G.,  2015, \mn@doi
  [\aap] {10.1051/0004-6361/201425481}, \href
  {https://ui.adsabs.harvard.edu/abs/2015A%26A...577A..42B} {577, A42}

\bibitem[\protect\citeauthoryear{{Bate}, {Bonnell}  \& {Bromm}}{{Bate}
  et~al.}{2002}]{Bate2002}
{Bate} M.~R.,  {Bonnell} I.~A.,   {Bromm} V.,  2002, \mn@doi [\mnras]
  {10.1046/j.1365-8711.2002.05539.x}, \href
  {https://ui.adsabs.harvard.edu/abs/2002MNRAS.332L..65B} {332, L65}

\bibitem[\protect\citeauthoryear{{Bayliss} et~al.,}{{Bayliss}
  et~al.}{2018}]{Bayliss2018}
{Bayliss} D.,  et~al., 2018, \mn@doi [\mnras] {10.1093/mnras/stx2778}, \href
  {http://adsabs.harvard.edu/abs/2018MNRAS.475.4467B} {475, 4467}

\bibitem[\protect\citeauthoryear{{Blake}, {Torres}, {Bloom}  \&
  {Gaudi}}{{Blake} et~al.}{2008}]{Blake2008}
{Blake} C.~H.,  {Torres} G.,  {Bloom} J.~S.,   {Gaudi} B.~S.,  2008, \mn@doi
  [\apj] {10.1086/589630}, \href
  {https://ui.adsabs.harvard.edu/abs/2008ApJ...684..635B} {684, 635}

\bibitem[\protect\citeauthoryear{{Blanco-Cuaresma}, {Soubiran}, {Heiter}  \&
  {Jofr{\'e}}}{{Blanco-Cuaresma} et~al.}{2014}]{Blanco2014}
{Blanco-Cuaresma} S.,  {Soubiran} C.,  {Heiter} U.,   {Jofr{\'e}} P.,  2014,
  \mn@doi [\aap] {10.1051/0004-6361/201423945}, \href
  {https://ui.adsabs.harvard.edu/abs/2014A&A...569A.111B} {569, A111}

\bibitem[\protect\citeauthoryear{{Bouchy} et~al.,}{{Bouchy}
  et~al.}{2011}]{Bouchy2011}
{Bouchy} F.,  et~al., 2011, \mn@doi [\aap] {10.1051/0004-6361/201117095}, \href
  {https://ui.adsabs.harvard.edu/abs/2011A&A...533A..83B} {533, A83}

\bibitem[\protect\citeauthoryear{{Casewell} et~al.,}{{Casewell}
  et~al.}{2018}]{Casewell2018}
{Casewell} S.~L.,  et~al., 2018, \mn@doi [\mnras] {10.1093/mnras/sty2183},
  \href {http://adsabs.harvard.edu/abs/2018MNRAS.481.1897C} {481, 1897}

\bibitem[\protect\citeauthoryear{{Castelli} \& {Kurucz}}{{Castelli} \&
  {Kurucz}}{2004}]{Castelli2004}
{Castelli} F.,  {Kurucz} R.~L.,  2004, preprint, \href
  {http://adsabs.harvard.edu/abs/2004astro.ph..5087C} {} (\mn@eprint {arXiv}
  {0405087})

\bibitem[\protect\citeauthoryear{{Chabrier}, {Gallardo}  \&
  {Baraffe}}{{Chabrier} et~al.}{2007}]{Chabrier2007}
{Chabrier} G.,  {Gallardo} J.,   {Baraffe} I.,  2007, \mn@doi [\aap]
  {10.1051/0004-6361:20077702}, \href
  {https://ui.adsabs.harvard.edu/abs/2007A&A...472L..17C} {472, L17}

\bibitem[\protect\citeauthoryear{{Choi}, {Dotter}, {Conroy}, {Cantiello},
  {Paxton}  \& {Johnson}}{{Choi} et~al.}{2016}]{Choi2016}
{Choi} J.,  {Dotter} A.,  {Conroy} C.,  {Cantiello} M.,  {Paxton} B.,
  {Johnson} B.~D.,  2016, \mn@doi [\apj] {10.3847/0004-637X/823/2/102}, \href
  {https://ui.adsabs.harvard.edu/abs/2016ApJ...823..102C} {823, 102}

\bibitem[\protect\citeauthoryear{{Coppejans} et~al.,}{{Coppejans}
  et~al.}{2013}]{Coppejans2013}
{Coppejans} R.,  et~al., 2013, \mn@doi [\pasp] {10.1086/672156}, \href
  {https://ui.adsabs.harvard.edu/abs/2013PASP..125..976C} {125, 976}

\bibitem[\protect\citeauthoryear{{Covey} et~al.,}{{Covey}
  et~al.}{2007}]{Covey2007}
{Covey} K.~R.,  et~al., 2007, \mn@doi [\aj] {10.1086/522052}, \href
  {https://ui.adsabs.harvard.edu/abs/2007AJ....134.2398C} {134, 2398}

\bibitem[\protect\citeauthoryear{{David}, {Hillenbrand}, {Gillen}, {Cody},
  {Howell}, {Isaacson}  \& {Livingston}}{{David} et~al.}{2019}]{David2019}
{David} T.~J.,  {Hillenbrand} L.~A.,  {Gillen} E.,  {Cody} A.~M.,  {Howell}
  S.~B.,  {Isaacson} H.~T.,   {Livingston} J.~H.,  2019, \mn@doi [\apj]
  {10.3847/1538-4357/aafe09}, \href
  {https://ui.adsabs.harvard.edu/abs/2019ApJ...872..161D} {872, 161}

\bibitem[\protect\citeauthoryear{{Delfosse} et~al.,}{{Delfosse}
  et~al.}{2004}]{Delfosse2004}
{Delfosse} X.,  et~al., 2004, {M dwarfs binaries: Results from accurate radial
  velocities and high angular resolution observations}.
pp 166--174

\bibitem[\protect\citeauthoryear{{Delrez} et~al.,}{{Delrez}
  et~al.}{2018}]{Delrez2018}
{Delrez} L.,  et~al., 2018, in \procspie. p. 107001I (\mn@eprint {arXiv}
  {1806.11205}), \mn@doi{10.1117/12.2312475}

\bibitem[\protect\citeauthoryear{{Dieterich}, {Henry}, {Jao}, {Winters},
  {Hosey}, {Riedel}  \& {Subasavage}}{{Dieterich} et~al.}{2014}]{Dieterich2014}
{Dieterich} S.~B.,  {Henry} T.~J.,  {Jao} W.-C.,  {Winters} J.~G.,  {Hosey}
  A.~D.,  {Riedel} A.~R.,   {Subasavage} J.~P.,  2014, \mn@doi [\aj]
  {10.1088/0004-6256/147/5/94}, \href
  {https://ui.adsabs.harvard.edu/abs/2014AJ....147...94D} {147, 94}

\bibitem[\protect\citeauthoryear{{Doyle} et~al.,}{{Doyle}
  et~al.}{2011}]{Doyle2011}
{Doyle} L.~R.,  et~al., 2011, \mn@doi [Science] {10.1126/science.1210923},
  \href {https://ui.adsabs.harvard.edu/abs/2011Sci...333.1602D} {333, 1602}

\bibitem[\protect\citeauthoryear{{Feiden} \& {Chaboyer}}{{Feiden} \&
  {Chaboyer}}{2012}]{Feiden2012}
{Feiden} G.~A.,  {Chaboyer} B.,  2012, \mn@doi [\apj]
  {10.1088/0004-637X/757/1/42}, \href
  {http://adsabs.harvard.edu/abs/2012ApJ...757...42F} {757, 42}

\bibitem[\protect\citeauthoryear{{Foreman-Mackey}, {Hogg}, {Lang}  \&
  {Goodman}}{{Foreman-Mackey} et~al.}{2013}]{Foreman-Mackey2013}
{Foreman-Mackey} D.,  {Hogg} D.~W.,  {Lang} D.,   {Goodman} J.,  2013, \mn@doi
  [\pasp] {10.1086/670067}, \href
  {http://adsabs.harvard.edu/abs/2013PASP..125..306F} {125, 306}

\bibitem[\protect\citeauthoryear{{Gaia Collaboration} et~al.,}{{Gaia
  Collaboration} et~al.}{2018}]{GAIA2018}
{Gaia Collaboration} et~al., 2018, \mn@doi [\aap]
  {10.1051/0004-6361/201833051}, \href
  {https://ui.adsabs.harvard.edu/abs/2018A&A...616A...1G} {616, A1}

\bibitem[\protect\citeauthoryear{{Gill} et~al.,}{{Gill}
  et~al.}{2019a}]{Gill2019a}
{Gill} S.,  et~al., 2019a, \mn@doi [\mnras] {10.1093/mnras/stz3212}, \href
  {https://ui.adsabs.harvard.edu/abs/2019MNRAS.tmp.2805G} {p.~2805}

\bibitem[\protect\citeauthoryear{Gill et~al.,}{Gill et~al.}{2019b}]{Gill2019b}
Gill S.,  et~al., 2019b, \mn@doi [Astronomy & Astrophysics]
  {10.1051/0004-6361/201833054}, 626, A119

\bibitem[\protect\citeauthoryear{{Gill} et~al.,}{{Gill}
  et~al.}{2020}]{Gill2020}
{Gill} S.,  et~al., 2020, \mn@doi [\apjl] {10.3847/2041-8213/ab9eb9}, \href
  {https://ui.adsabs.harvard.edu/abs/2020ApJ...898L..11G} {898, L11}

\bibitem[\protect\citeauthoryear{{Gillen}, {Hillenbrand}, {David}, {Aigrain},
  {Rebull}, {Stauffer}, {Cody}  \& {Queloz}}{{Gillen}
  et~al.}{2017}]{Gillen2017}
{Gillen} E.,  {Hillenbrand} L.~A.,  {David} T.~J.,  {Aigrain} S.,  {Rebull} L.,
   {Stauffer} J.,  {Cody} A.~M.,   {Queloz} D.,  2017, \mn@doi [\apj]
  {10.3847/1538-4357/aa84b3}, \href
  {https://ui.adsabs.harvard.edu/abs/2017ApJ...849...11G} {849, 11}

\bibitem[\protect\citeauthoryear{{Gillon}, {Jehin}, {Magain}, {Chantry},
  {Hutsem{\'e}kers}, {Manfroid}, {Queloz}  \& {Udry}}{{Gillon}
  et~al.}{2011}]{Gillon2011}
{Gillon} M.,  {Jehin} E.,  {Magain} P.,  {Chantry} V.,  {Hutsem{\'e}kers} D.,
  {Manfroid} J.,  {Queloz} D.,   {Udry} S.,  2011, in European Physical Journal
  Web of Conferences. p. 06002 (\mn@eprint {arXiv} {1101.5807}),
  \mn@doi{10.1051/epjconf/20101106002}

\bibitem[\protect\citeauthoryear{{Gillon} et~al.,}{{Gillon}
  et~al.}{2017}]{Gillon2017}
{Gillon} M.,  et~al., 2017, \mn@doi [\nat] {10.1038/nature21360}, \href
  {http://adsabs.harvard.edu/abs/2017Natur.542..456G} {542, 456}

\bibitem[\protect\citeauthoryear{{Goodwin} \& {Whitworth}}{{Goodwin} \&
  {Whitworth}}{2007}]{Goodwin2007}
{Goodwin} S.~P.,  {Whitworth} A.,  2007, \mn@doi [\aap]
  {10.1051/0004-6361:20066745}, \href
  {https://ui.adsabs.harvard.edu/abs/2007A&A...466..943G} {466, 943}

\bibitem[\protect\citeauthoryear{{G{\"u}nther} et~al.,}{{G{\"u}nther}
  et~al.}{2017}]{Gunther2017}
{G{\"u}nther} M.~N.,  et~al., 2017, \mn@doi [\mnras] {10.1093/mnras/stx1920},
  \href {https://ui.adsabs.harvard.edu/abs/2017MNRAS.472..295G} {472, 295}

\bibitem[\protect\citeauthoryear{{G{\"u}nther} et~al.,}{{G{\"u}nther}
  et~al.}{2019}]{Guenther2019}
{G{\"u}nther} M.~N.,  et~al., 2019, \mn@doi [Nature Astronomy]
  {10.1038/s41550-019-0845-5}, \href
  {https://ui.adsabs.harvard.edu/abs/2019NatAs.tmp..420G} {p.~420}

\bibitem[\protect\citeauthoryear{Gómez Maqueo~Chew et~al.,}{Gómez Maqueo~Chew
  et~al.}{2014}]{Chew2014}
Gómez Maqueo~Chew Y.,  et~al., 2014, \mn@doi [Astronomy & Astrophysics]
  {10.1051/0004-6361/201424265}, 572, A50

\bibitem[\protect\citeauthoryear{{Hauschildt}, {Allard}  \&
  {Baron}}{{Hauschildt} et~al.}{1999}]{Hauschildt99}
{Hauschildt} P.~H.,  {Allard} F.,   {Baron} E.,  1999, \mn@doi [\apj]
  {10.1086/306745}, \href {http://adsabs.harvard.edu/abs/1999ApJ...512..377H}
  {512, 377}

\bibitem[\protect\citeauthoryear{{Henry}, {Jao}, {Subasavage}, {Beaulieu},
  {Ianna}, {Costa}  \& {M{\'e}ndez}}{{Henry} et~al.}{2006}]{Henry2006}
{Henry} T.~J.,  {Jao} W.-C.,  {Subasavage} J.~P.,  {Beaulieu} T.~D.,  {Ianna}
  P.~A.,  {Costa} E.,   {M{\'e}ndez} R.~A.,  2006, \mn@doi [\aj]
  {10.1086/508233}, \href
  {https://ui.adsabs.harvard.edu/abs/2006AJ....132.2360H} {132, 2360}

\bibitem[\protect\citeauthoryear{Husser, {Wende-von Berg}, Dreizler, Homeier,
  Reiners, Barman  \& Hauschildt}{Husser et~al.}{2013}]{Husser2013}
Husser T.-O.,  {Wende-von Berg} S.,  Dreizler S.,  Homeier D.,  Reiners A.,
  Barman T.,   Hauschildt P.~H.,  2013, \mn@doi [A{\&}A]
  {10.1051/0004-6361/201219058}, 553, A6

\bibitem[\protect\citeauthoryear{{Irwin} et~al.,}{{Irwin}
  et~al.}{2010}]{Irwin2010}
{Irwin} J.,  et~al., 2010, \mn@doi [\apj] {10.1088/0004-637X/718/2/1353}, \href
  {https://ui.adsabs.harvard.edu/abs/2010ApJ...718.1353I} {718, 1353}

\bibitem[\protect\citeauthoryear{{Irwin} et~al.,}{{Irwin}
  et~al.}{2011}]{Irwin2011}
{Irwin} J.~M.,  et~al., 2011, \mn@doi [\apj] {10.1088/0004-637X/742/2/123},
  \href {https://ui.adsabs.harvard.edu/abs/2011ApJ...742..123I} {742, 123}

\bibitem[\protect\citeauthoryear{{Jackman} et~al.,}{{Jackman}
  et~al.}{2019}]{Jackman2019}
{Jackman} J. A.~G.,  et~al., 2019, \mn@doi [\mnras] {10.1093/mnras/stz2496},
  \href {https://ui.adsabs.harvard.edu/abs/2019MNRAS.489.5146J} {489, 5146}

\bibitem[\protect\citeauthoryear{{Johnson} et~al.,}{{Johnson}
  et~al.}{2011}]{Johnson2011}
{Johnson} J.~A.,  et~al., 2011, \mn@doi [\apj] {10.1088/0004-637X/730/2/79},
  \href {https://ui.adsabs.harvard.edu/abs/2011ApJ...730...79J} {730, 79}

\bibitem[\protect\citeauthoryear{Kaltenegger \& Traub}{Kaltenegger \&
  Traub}{2009}]{Kaltenegger2009}
Kaltenegger L.,  Traub W.~A.,  2009, \mn@doi [The Astrophysical Journal]
  {10.1088/0004-637x/698/1/519}, 698, 519–527

\bibitem[\protect\citeauthoryear{{Kesseli}, {West}, {Veyette}, {Harrison},
  {Feldman}  \& {Bochanski}}{{Kesseli} et~al.}{2017}]{Kesseli2017}
{Kesseli} A.~Y.,  {West} A.~A.,  {Veyette} M.,  {Harrison} B.,  {Feldman} D.,
  {Bochanski} J.~J.,  2017, \mn@doi [\apjs] {10.3847/1538-4365/aa656d}, \href
  {https://ui.adsabs.harvard.edu/abs/2017ApJS..230...16K} {230, 16}

\bibitem[\protect\citeauthoryear{{Kostov} et~al.,}{{Kostov}
  et~al.}{2019}]{Kostov2019}
{Kostov} V.~B.,  et~al., 2019, arXiv e-prints, \href
  {http://adsabs.harvard.edu/abs/2019arXiv190308017K} {}

\bibitem[\protect\citeauthoryear{{Kov{\'a}cs}, {Zucker}  \&
  {Mazeh}}{{Kov{\'a}cs} et~al.}{2016}]{Kovacs2016}
{Kov{\'a}cs} G.,  {Zucker} S.,   {Mazeh} T.,  2016, {BLS: Box-fitting Least
  Squares}, Astrophysics Source Code Library (\mn@eprint {ascl} {1607.008})

\bibitem[\protect\citeauthoryear{{Kraus}, {Tucker}, {Thompson}, {Craine}  \&
  {Hillenbrand}}{{Kraus} et~al.}{2011}]{Kraus2011}
{Kraus} A.~L.,  {Tucker} R.~A.,  {Thompson} M.~I.,  {Craine} E.~R.,
  {Hillenbrand} L.~A.,  2011, \mn@doi [\apj] {10.1088/0004-637X/728/1/48},
  \href {https://ui.adsabs.harvard.edu/abs/2011ApJ...728...48K} {728, 48}

\bibitem[\protect\citeauthoryear{{Kurucz}}{{Kurucz}}{1993}]{Kurucz1993}
{Kurucz} R.~L.,  1993, VizieR Online Data Catalog, \href
  {http://adsabs.harvard.edu/abs/1993yCat.6039....0K} {6039}

\bibitem[\protect\citeauthoryear{{Laithwaite} \& {Warren}}{{Laithwaite} \&
  {Warren}}{2020}]{Laithwaite2020}
{Laithwaite} R.~C.,  {Warren} S.~J.,  2020, arXiv e-prints, \href
  {https://ui.adsabs.harvard.edu/abs/2020arXiv200611092L} {p. arXiv:2006.11092}

\bibitem[\protect\citeauthoryear{{Lendl} et~al.,}{{Lendl}
  et~al.}{2019}]{Lendl2019}
{Lendl} M.,  et~al., 2019, \mn@doi [\mnras] {10.1093/mnras/stz3545}, \href
  {https://ui.adsabs.harvard.edu/abs/2019MNRAS.tmp.3189L} {p.~3189}

\bibitem[\protect\citeauthoryear{{L{\'o}pez-Morales}}{{L{\'o}pez-Morales}}{2007}]{Lopez2007}
{L{\'o}pez-Morales} M.,  2007, \mn@doi [\apj] {10.1086/513142}, \href
  {https://ui.adsabs.harvard.edu/abs/2007ApJ...660..732L} {660, 732}

\bibitem[\protect\citeauthoryear{{Lucy} \& {Sweeney}}{{Lucy} \&
  {Sweeney}}{1971}]{Lucy1971}
{Lucy} L.~B.,  {Sweeney} M.~A.,  1971, \mn@doi [\aj] {10.1086/111159}, \href
  {https://ui.adsabs.harvard.edu/abs/1971AJ.....76..544L} {76, 544}

\bibitem[\protect\citeauthoryear{{Maxted}}{{Maxted}}{2016}]{Maxted2016}
{Maxted} P.~F.~L.,  2016, \mn@doi [\aap] {10.1051/0004-6361/201628579}, \href
  {http://adsabs.harvard.edu/abs/2016A%26A...591A.111M} {591, A111}

\bibitem[\protect\citeauthoryear{{Mayor} et~al.,}{{Mayor}
  et~al.}{2003}]{Mayor2003}
{Mayor} M.,  et~al., 2003, The Messenger, \href
  {http://adsabs.harvard.edu/abs/2003Msngr.114...20M} {114, 20}

\bibitem[\protect\citeauthoryear{{Mireles} et~al.,}{{Mireles}
  et~al.}{2020}]{Mireles2020}
{Mireles} I.,  et~al., 2020, arXiv e-prints, \href
  {https://ui.adsabs.harvard.edu/abs/2020arXiv200614019M} {p. arXiv:2006.14019}

\bibitem[\protect\citeauthoryear{{Montet} et~al.,}{{Montet}
  et~al.}{2015}]{Montet2015}
{Montet} B.~T.,  et~al., 2015, \mn@doi [\apj] {10.1088/0004-637X/800/2/134},
  \href {https://ui.adsabs.harvard.edu/abs/2015ApJ...800..134M} {800, 134}

\bibitem[\protect\citeauthoryear{Nefs et~al.,}{Nefs et~al.}{2013}]{Nefs2013}
Nefs S.~V.,  et~al., 2013, \mn@doi [Monthly Notices of the Royal Astronomical
  Society] {10.1093/mnras/stt405}, 431, 3240

\bibitem[\protect\citeauthoryear{{Parsons} et~al.,}{{Parsons}
  et~al.}{2018}]{Parsons2018}
{Parsons} S.~G.,  et~al., 2018, \mn@doi [\mnras] {10.1093/mnras/sty2345}, \href
  {https://ui.adsabs.harvard.edu/abs/2018MNRAS.481.1083P} {481, 1083}

\bibitem[\protect\citeauthoryear{{Parviainen} \& {Aigrain}}{{Parviainen} \&
  {Aigrain}}{2015}]{Parvianen2015}
{Parviainen} H.,  {Aigrain} S.,  2015, \mn@doi [\mnras]
  {10.1093/mnras/stv1857}, \href
  {http://adsabs.harvard.edu/abs/2015MNRAS.453.3821P} {453, 3821}

\bibitem[\protect\citeauthoryear{{Ribas}}{{Ribas}}{2006}]{Ribas2006}
{Ribas} I.,  2006, \mn@doi [\apss] {10.1007/s10509-006-9081-4}, \href
  {https://ui.adsabs.harvard.edu/abs/2006Ap&SS.304...89R} {304, 89}

\bibitem[\protect\citeauthoryear{{Ricker} et~al.,}{{Ricker}
  et~al.}{2014}]{Ricker2014}
{Ricker} G.~R.,  et~al., 2014, in \procspie. p. 914320 (\mn@eprint {arXiv}
  {1406.0151}), \mn@doi{10.1117/12.2063489}

\bibitem[\protect\citeauthoryear{Schlafly \& Finkbeiner}{Schlafly \&
  Finkbeiner}{2011}]{Schlafly2011}
Schlafly E.~F.,  Finkbeiner D.~P.,  2011, \mn@doi [Astrophysical Journal]
  {10.1088/0004-637X/737/2/103}, 737

\bibitem[\protect\citeauthoryear{{Schlegel}, {Finkbeiner}  \&
  {Davis}}{{Schlegel} et~al.}{1998}]{Schlegel1998}
{Schlegel} D.~J.,  {Finkbeiner} D.~P.,   {Davis} M.,  1998, \mn@doi [\apj]
  {10.1086/305772}, \href
  {https://ui.adsabs.harvard.edu/abs/1998ApJ...500..525S} {500, 525}

\bibitem[\protect\citeauthoryear{{Skrutskie} et~al.,}{{Skrutskie}
  et~al.}{2006}]{Skrutskie2006}
{Skrutskie} M.~F.,  et~al., 2006, \mn@doi [\aj] {10.1086/498708}, \href
  {https://ui.adsabs.harvard.edu/abs/2006AJ....131.1163S} {131, 1163}

\bibitem[\protect\citeauthoryear{{Southworth} et~al.,}{{Southworth}
  et~al.}{2015}]{Southworth2015}
{Southworth} J.,  et~al., 2015, \mn@doi [\mnras] {10.1093/mnras/stv2183}, \href
  {https://ui.adsabs.harvard.edu/abs/2015MNRAS.454.3094S} {454, 3094}

\bibitem[\protect\citeauthoryear{{Spada}, {Demarque}, {Kim}  \&
  {Sills}}{{Spada} et~al.}{2013}]{Spada2013}
{Spada} F.,  {Demarque} P.,  {Kim} Y.~C.,   {Sills} A.,  2013, \mn@doi [\apj]
  {10.1088/0004-637X/776/2/87}, \href
  {https://ui.adsabs.harvard.edu/abs/2013ApJ...776...87S} {776, 87}

\bibitem[\protect\citeauthoryear{{Speagle}}{{Speagle}}{2019}]{Speagle2019}
{Speagle} J.~S.,  2019, arXiv e-prints, \href
  {https://ui.adsabs.harvard.edu/abs/2019arXiv190402180S} {p. arXiv:1904.02180}

\bibitem[\protect\citeauthoryear{{Stassun} et~al.,}{{Stassun}
  et~al.}{2019}]{Stassun2019}
{Stassun} K.~G.,  et~al., 2019, \mn@doi [\aj] {10.3847/1538-3881/ab3467}, \href
  {https://ui.adsabs.harvard.edu/abs/2019AJ....158..138S} {158, 138}

\bibitem[\protect\citeauthoryear{{Stevens} et~al.,}{{Stevens}
  et~al.}{2019}]{Stevens2019}
{Stevens} D.~J.,  et~al., 2019, arXiv e-prints, \href
  {https://ui.adsabs.harvard.edu/abs/2019arXiv191006212S} {p. arXiv:1910.06212}

\bibitem[\protect\citeauthoryear{{Tamuz}, {Mazeh}  \& {Zucker}}{{Tamuz}
  et~al.}{2005}]{2005MNRAS.356.1466T}
{Tamuz} O.,  {Mazeh} T.,   {Zucker} S.,  2005, \mn@doi [\mnras]
  {10.1111/j.1365-2966.2004.08585.x}, \href
  {https://ui.adsabs.harvard.edu/abs/2005MNRAS.356.1466T} {356, 1466}

\bibitem[\protect\citeauthoryear{{Terrien}, {Fleming}, {Mahadevan},
  {Deshpande}, {Feiden}, {Bender}  \& {Ramsey}}{{Terrien}
  et~al.}{2012}]{Terrien2012}
{Terrien} R.~C.,  {Fleming} S.~W.,  {Mahadevan} S.,  {Deshpande} R.,  {Feiden}
  G.~A.,  {Bender} C.~F.,   {Ramsey} L.~W.,  2012, \mn@doi [\apjl]
  {10.1088/2041-8205/760/1/L9}, \href
  {http://adsabs.harvard.edu/abs/2012ApJ...760L...9T} {760, L9}

\bibitem[\protect\citeauthoryear{Triaud et~al.,}{Triaud
  et~al.}{2012}]{Triaud2012}
Triaud A. H. M.~J.,  et~al., 2012, \mn@doi [Astronomy & Astrophysics]
  {10.1051/0004-6361/201219643}, 549, A18

\bibitem[\protect\citeauthoryear{Triaud et~al.,}{Triaud
  et~al.}{2017}]{Triaud2017}
Triaud A. H. M.~J.,  et~al., 2017, \mn@doi [Astronomy & Astrophysics]
  {10.1051/0004-6361/201730993}, 608, A129

\bibitem[\protect\citeauthoryear{{Triaud} et~al.,}{{Triaud}
  et~al.}{2020}]{Triaud2020b}
{Triaud} A. H.~M.~J.,  et~al., 2020, \mn@doi [Nature Astronomy]
  {10.1038/s41550-020-1018-2}, \href
  {https://ui.adsabs.harvard.edu/abs/2020NatAs...4..650T} {4, 650}

\bibitem[\protect\citeauthoryear{{Vines} \& {Jenkins}}{{Vines} \&
  {Jenkins}}{2020}]{Vines2020}
{Vines} J.~I.,  {Jenkins} J.~S.,  2020, in prep

\bibitem[\protect\citeauthoryear{{Wheatley} et~al.,}{{Wheatley}
  et~al.}{2018}]{Wheatley2018}
{Wheatley} P.~J.,  et~al., 2018, \mn@doi [\mnras] {10.1093/mnras/stx2836},
  \href {http://adsabs.harvard.edu/abs/2018MNRAS.475.4476W} {475, 4476}

\bibitem[\protect\citeauthoryear{{Wisniewski} et~al.,}{{Wisniewski}
  et~al.}{2012}]{Wisniewski2012}
{Wisniewski} J.~P.,  et~al., 2012, \mn@doi [\aj] {10.1088/0004-6256/143/5/107},
  \href {https://ui.adsabs.harvard.edu/abs/2012AJ....143..107W} {143, 107}

\bibitem[\protect\citeauthoryear{{Yee}, {Petigura}  \& {von Braun}}{{Yee}
  et~al.}{2017}]{Yee2017}
{Yee} S.~W.,  {Petigura} E.~A.,   {von Braun} K.,  2017, \mn@doi [\apj]
  {10.3847/1538-4357/836/1/77}, \href
  {https://ui.adsabs.harvard.edu/abs/2017ApJ...836...77Y} {836, 77}

\bibitem[\protect\citeauthoryear{von Boetticher et~al.,}{von Boetticher
  et~al.}{2017}]{von_Boetticher2017}
von Boetticher A.,  et~al., 2017, \mn@doi [Astronomy & Astrophysics]
  {10.1051/0004-6361/201731107}, 604, L6

\bibitem[\protect\citeauthoryear{von Boetticher et~al.,}{von Boetticher
  et~al.}{2019}]{vonBoetticher_2019}
von Boetticher A.,  et~al., 2019, \mn@doi [Astronomy & Astrophysics]
  {10.1051/0004-6361/201834539}, 625, A150

\makeatother
\end{thebibliography}

\bsp	
\label{lastpage}
\end{document}